\documentclass[sigconf, screen]{acmart}
\usepackage{amsmath}
\usepackage{url}
\usepackage{subfigure}
\usepackage{graphicx}
\usepackage{mdframed}
\usepackage{enumitem}
\usepackage[linesnumbered,ruled,vlined,lined,boxed,commentsnumbered]{algorithm2e}
\newmdenv[
    backgroundcolor=gray!20,
    linecolor=black,
    linewidth=1pt,
    roundcorner=5pt,
    frametitle={Prompt},
    frametitlefont=\bfseries,
    frametitlebackgroundcolor=gray!30,
]{promptbox}
\usepackage{balance}
\usepackage{hyperref}
\hypersetup{hidelinks,
	colorlinks=true,
	allcolors=black,
	pdfstartview=Fit,
	breaklinks=true}

\newcommand{\tealcomment}[1]{{\color{teal}{// #1}}}
\copyrightyear{2025}
\acmYear{2025}
\setcopyright{cc}
\setcctype{by}
\acmConference[SIGCOMM '25]{ACM SIGCOMM 2025 Conference}{September 8--11, 2025}{Coimbra, Portugal}
\acmBooktitle{ACM SIGCOMM 2025 Conference (SIGCOMM '25), September 8--11, 2025, Coimbra, Portugal}
\acmDOI{10.1145/3718958.3750526}
\acmISBN{979-8-4007-1524-2/2025/09}
\copyrightyear{2025}
\acmYear{2025}

\begin{document}

\title{Towards User-level QoE: Large-scale Practice in Personalized Optimization of Adaptive Video Streaming
}

\author{Lianchen Jia$^{1,3}$, Chao Zhou$^{2*}$, Chaoyang Li$^{1}$, Jiangchuan Liu$^{3}$, Lifeng Sun$^{1,4}$}

\authornote{Chao Zhou and Lifeng Sun are corresponding authors.}
\affiliation{
  $^{1}$Department of Computer Science and Technology, Tsinghua University\country{China},
  $^{2}$Kuaishou\country{China},
  $^{3}$Simon Fraser University\country{Canada},
  $^{4}$BNRist\country{China},
}

\email{jlc21@mails.tsinghua.edu.cn, zhouchaoyf@gmail.com, sunlf@tsinghua.edu.cn}

\renewcommand{\shortauthors}{Lianchen Jia, Chao Zhou, Chaoyang Li, Jiangchuan Liu, Lifeng Sun}

\begin{abstract}
Traditional optimization methods based on system-wide Quality of Service (QoS) metrics have approached their performance limitations in modern large-scale streaming systems.  However, aligning user-level Quality of Experience~(QoE) with algorithmic optimization objectives remains an unresolved challenge. Therefore, we propose \texttt{LingXi}, the first large-scale deployed system for personalized adaptive video streaming based on user-level experience. \texttt{LingXi} dynamically optimizes the objectives of adaptive video streaming algorithms by analyzing user engagement. Utilizing exit rate as a key metric, we investigate the correlation between QoS indicators and exit rates based on production environment logs, subsequently developing a personalized exit rate predictor. Through Monte Carlo sampling and online Bayesian optimization, we iteratively determine optimal parameters. Large-scale A/B testing utilizing 8\% of traffic on Kuaishou, one of the largest short video platforms, demonstrates \texttt{LingXi}'s superior performance.  \texttt{LingXi} achieves a 0.15\% increase in total viewing time, a 0.1\% improvement in bitrate, and a 1.3\% reduction in stall time across all users, with particularly significant improvements for low-bandwidth users who experience a 15\% reduction in stall time.
\end{abstract}

\maketitle

\section{INTRODUCTION}
With the advancement of network technologies (such as 5G and WiFi6)~\cite{narayanan20215g,wifi6} and the growing demand for user-generated content~\cite{naab2017studies}, video has become an indispensable part of people's daily lives. According to the 2024 Global Internet Phenomena Report~\cite{sandvine24}, video applications are the primary contributors to downstream traffic, accounting for 39\% and 31\% of traffic in fixed and mobile networks, respectively.  On short video platform Kuaishou, the average daily active users reach 400 million, with each daily active user spending an average of 125.6 minutes per day on the platform~\cite{kuaishou2025q4results}. Video streaming systems aim to enhance users' Quality of Experience (QoE), which the ITU-T standardization~\cite{itu1999subjective} defines as "the overall acceptability of an application or service, as perceived subjectively by the end-user". In production environments, user engagement metrics, including watch time and exit rate, serve as primary indicators of QoE~\cite{dobrian2011understanding}.

Adaptive Bitrate (ABR) algorithms play a crucial role in determining user-level QoE in network video streaming~\cite{seufert2014survey}. These algorithms dynamically optimize bitrate selection for upcoming video segments based on current buffer status and historical bandwidth measurements, striving to enhance video quality and playback smoothness while minimizing stalling events. Despite representing the ultimate optimization goal, QoE metrics pose significant challenges for direct optimization approaches~\cite{ghadiyaram2017subjective}. Over the past decade, researchers have primarily focused on approximating QoE through measurable Quality of Service (QoS) indicators~\cite{yin2015control,sqoe4} as proxy optimization targets. In algorithmic design, linear optimization objectives weighted by QoS indicators (denoted as $QoE_{lin}$) have emerged as the dominant approach~\cite{yin2015control,yan2020fugu}, establishing a standardized framework for algorithm development and evaluation.

\textit{\textbf{Golden Years, but it's time to blaze a new trail.}}
With recent advancements in video streaming technology~\cite{mao2017neural,alomar2022causalsim,yin2015control} and network infrastructure~\cite{narayanan20215g,wifi6}, traditional optimization objectives are rapidly nearing their performance ceiling. Our 5-day A/B testing demonstrates that while current QoS metrics and the linear user experience model $QoE_{lin}$ still show modest potential improvements of 0.5\% to 2\% in production systems, these metrics may counterintuitively impact core QoE indicators, such as watch time~(as discussed in \S\ref{When the garden is well-tended}). This limitation demands a paradigm shift: while system-level QoS optimization may have plateaued, substantial opportunities remain in optimizing algorithmic performance for individual user QoE. Research has demonstrated considerable variation in users' QoS sensitivity~\cite{sqoe4}. Consequently, unified $QoE_{lin}$\cite{yin2015control} formulations derived through offline analysis (or neural network-fitted $QoE_{nn}$\cite{sqoe4,huang2023optimizing} for Mean Opinion Scores (MOS)) can only effectively represent average user experiences, failing to capture individual-specific QoE characteristics.

\begin{table*}[h]
\centering
\caption{Differences between \texttt{LingXi} and Previous Works}
\vspace{-5pt}
\setlength{\tabcolsep}{0.5mm}
\begin{tabular}{l|c|c|c|c|c}
&No Explicit Feedback&No Negative Influence&Continuous Optimization&User-Level&Large-scale Deployed\\
\hline
Offline Data Analysis~\cite{sqoe4,huang2023optimizing}&$\checkmark$&$\checkmark$&&&$\checkmark$\\
\hline
Lab Study~\cite{zuo2022adaptive,qiao2020beyond}&&&&$\checkmark$&\\

\hline
Intervention  Playback~\cite{zhang2022enabling}&$\checkmark$&&$\checkmark$&$\checkmark$&\\
\hline
\textbf{\texttt{LingXi}~(Ours)}&\pmb{$\checkmark$}&\pmb{$\checkmark$}&\pmb{$\checkmark$}&\pmb{$\checkmark$}&\pmb{$\checkmark$}\\
\end{tabular}
\vspace{-10pt}
\label{tab:Differences from Previous Work}
\end{table*}
\textit{\textbf{The last piece of the puzzle.}}
While QoE personalization is not unprecedented, existing approaches have primarily targeted higher-layer components within the video streaming architecture, such as personalized content delivery scheduling~\cite{dobrian2011understanding,krishnan2012video,wang2019intelligent}. These methods leverage comprehensive user profiles constructed from extensive engagement data, typically requiring stable and substantial historical information. However, user-level personalization optimization has remained largely restricted to small-scale laboratory experiments.
Recent studies have explored personalized QoE through user-level models developed via subjective evaluation experiments~\cite{qiao2020beyond,zuo2022adaptive}. However, the requirement for explicit user feedback significantly compromises user experience, inherently limiting these methods' scalability and practical viability in large-scale deployments~\cite{zhao2016qoe}. Alternative approaches~\cite{zhang2022enabling} attempt to rapidly profile new users through playback interventions, such as bandwidth limitations, but these interventions inevitably degrade user experience. We summarize the key distinctions between \texttt{LingXi} and existing approaches in Table~\ref{tab:Differences from Previous Work}.

To complete the final piece of the puzzle, we propose \texttt{LingXi}, the first large-scale deployed adaptive video streaming system for user-level personalized QoE. \texttt{LingXi} optimizes ABR algorithm objectives through online user engagement perception during playback sessions. We utilize segment-level exit rate as a key metric to bridge the gap between optimization objectives and user engagement metrics. Through systematic analysis, we investigate the correlation between exit rate and three fundamental QoS metrics: video quality, video smoothness, and stall time. Our findings reveal that these factors influence exit rates at distinctly different orders of magnitude ~(\S\ref{Exit Rate: QoE Metric for Analysi}). Consequently, the personalized modeling of metrics with relatively minor effects (such as video quality and smoothness) can be masked by stochastic fluctuations~(evaluated in~\S\ref{sec:exit rate}). We then analyze users' perception of stall events, which have the most substantial impact on QoE, revealing both individual differences and temporal dynamics~(\S\ref{sec:Online Log Analysis for QoS}).

\texttt{LingXi} supports arbitrary ABR algorithms (regardless of whether they have explicit optimization objectives)\cite{huang2014buffer,yin2015control,mao2017neural,spiteri2020bola,akhtar2018oboe,yan2020fugu,jia2024dancing,jia2023rdladder} by incorporating a dynamic QoE adjustment module that modifies optimization objectives during runtime. The system tracks comprehensive state, including historical stall, user engagement, buffer occupancy, and bitrate. We implement dynamic parameter optimization through online Bayesian optimization\cite{williams2006gaussian} to address temporal dynamics in user perception~(\S\ref{sec:Bayesian Optimization}). Through Monte Carlo sampling~\cite{rubinstein2016simulation}, we simulate diverse playback scenarios in a virtual environment, generating playback traces under given parameters for predictor evaluation~(\S\ref{sec:Monte Carlo Alignment Perception}). To address the heterogeneity in QoS perception, we develop a hybrid framework that combines personalized neural network modeling with overall statistical analysis to create a robust exit rate predictor~(\S\ref{sec:Exit Rate Predictor}). Additionally, we detail our deployment architecture, including system integration, activation triggers, and optimization pruning mechanisms~(\S\ref{sec:Deployment}).

Comprehensive A/B testing of \texttt{LingXi} in production environment demonstrates significant performance improvements. Compared to the existing production algorithm, \texttt{LingXi} achieves a 0.15\% increase in total watch time, a 0.1\% enhancement in average bitrate, and a 1.3\% reduction in total stall time. \texttt{LingXi}'s performance in bandwidth-constrained conditions (below 2000 kbps)  achieves approximately 15\% reduction in stall time while maintaining comparable video quality. Detailed analysis of user-level adaptations reveals correlations between individual users' stall sensitivity and corresponding algorithmic parameter adjustments.

In summary, our key contributions are as follows:
\begin{itemize}[leftmargin=18pt]
\item Through online A/B testing, we discover that system-wide QoS optimization approaches are approaching saturation, necessitating a shift towards new optimization objectives. We conduct a comprehensive analysis of 1.5 million watch trajectories from production platforms to systematically evaluate the impact of various QoS metrics on QoE, revealing different QoS heterogeneous effects. Further investigation into stall events demonstrates both individual variations and temporal dynamics in QoE~(\S\ref{sec:QoS Impact on Exit Rate}).

\item We present \texttt{LingXi}, the first large-scale deployed system for user-level personalized adaptive video streaming. The system implements dynamic optimization objective adjustments based on historical user responses to stall events. We develop a hybrid exit rate model that integrates personalized neural network modeling with overall statistical modeling. We employ Monte Carlo sampling to generate QoS traces under various parameter settings for predictor evaluation while utilizing online Bayesian optimization for continuous parameter refinement~(\S\ref{sec:sys}).  Furthermore, we detail the integration methodology for existing ABR systems, encompassing integration approaches, triggering conditions, and pruning strategies~(\S\ref{sec:Deployment}).

\item Extensive A/B testing in production environment demonstrates \texttt{LingXi}'s capability to align with user-level QoE. Compared to the existing online algorithm, the system achieves a 0.15\% increase in total watch time, a 0.1\% enhancement in average bitrate, and a 1.3\% reduction in total stall time.~(\S\ref{sec:Evaluation}).
\end{itemize}

\begin{figure}
    \centering
      \subfigure[Video Bitrate: $Alg_3$ achieved the highest video quality]{
    \includegraphics[width=0.45\linewidth]{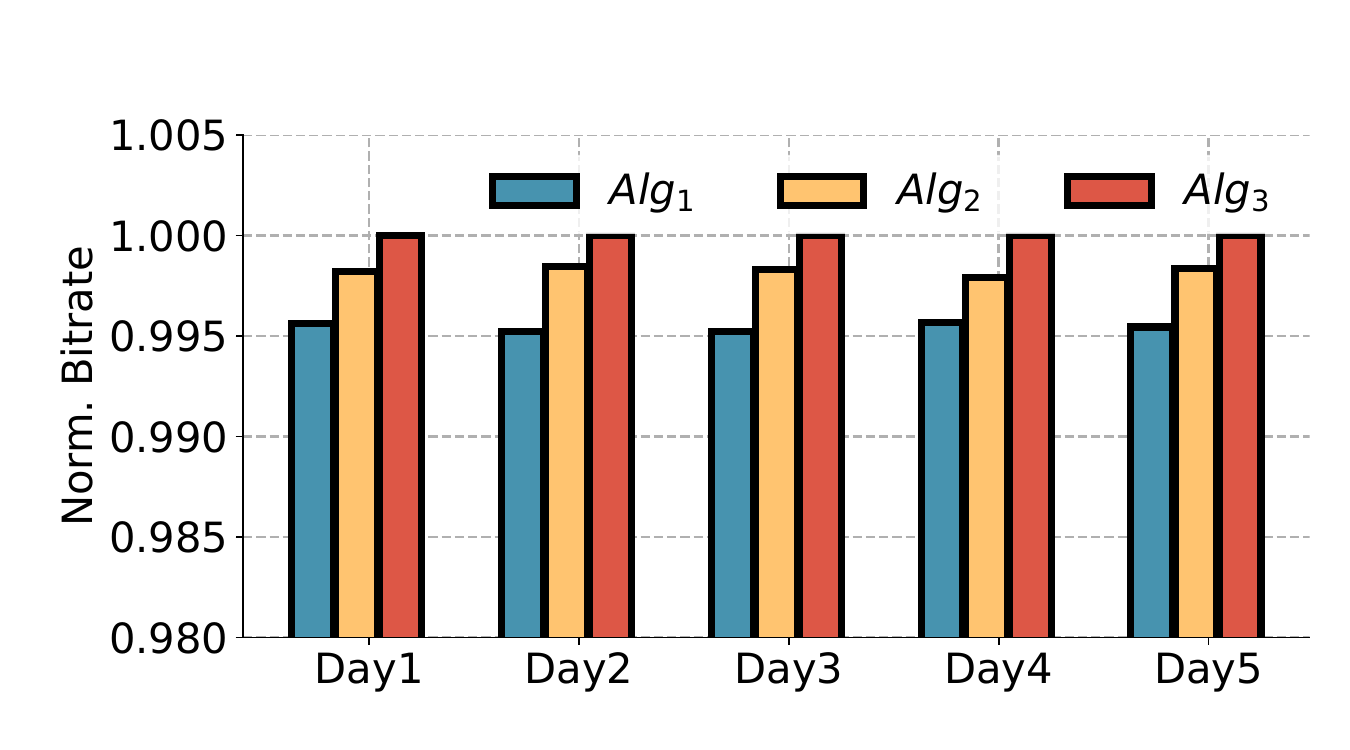}
    \label{img/intro/abnetnewq.pdf}
}
\subfigure[Stall Time: $Alg_1$ achieved the lowest stall time]{
    \includegraphics[width=0.45\linewidth]{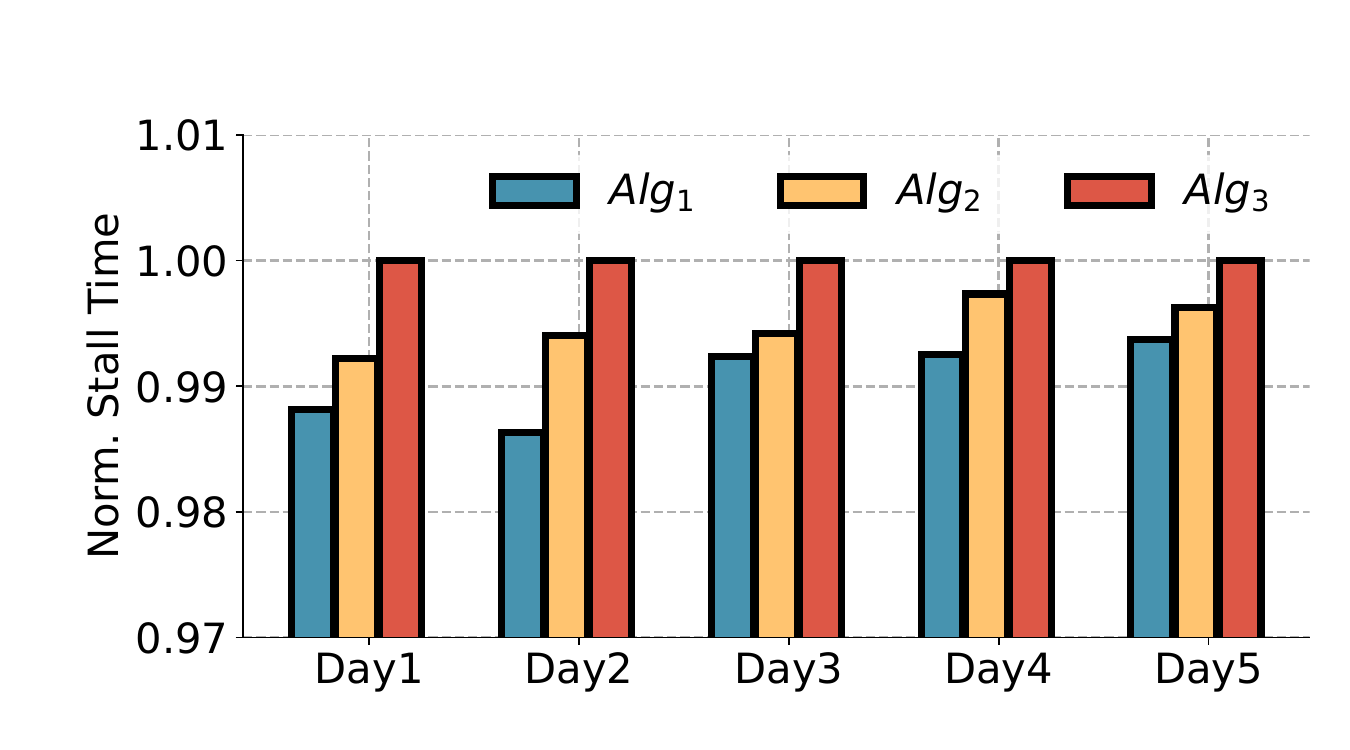}
    \label{img/mot/switch_1.pdf}
}
\subfigure[$QoE_{lin}$: $Alg_1$ obtained the highest $QoE_{lin}$]{
    \includegraphics[width=0.45\linewidth]{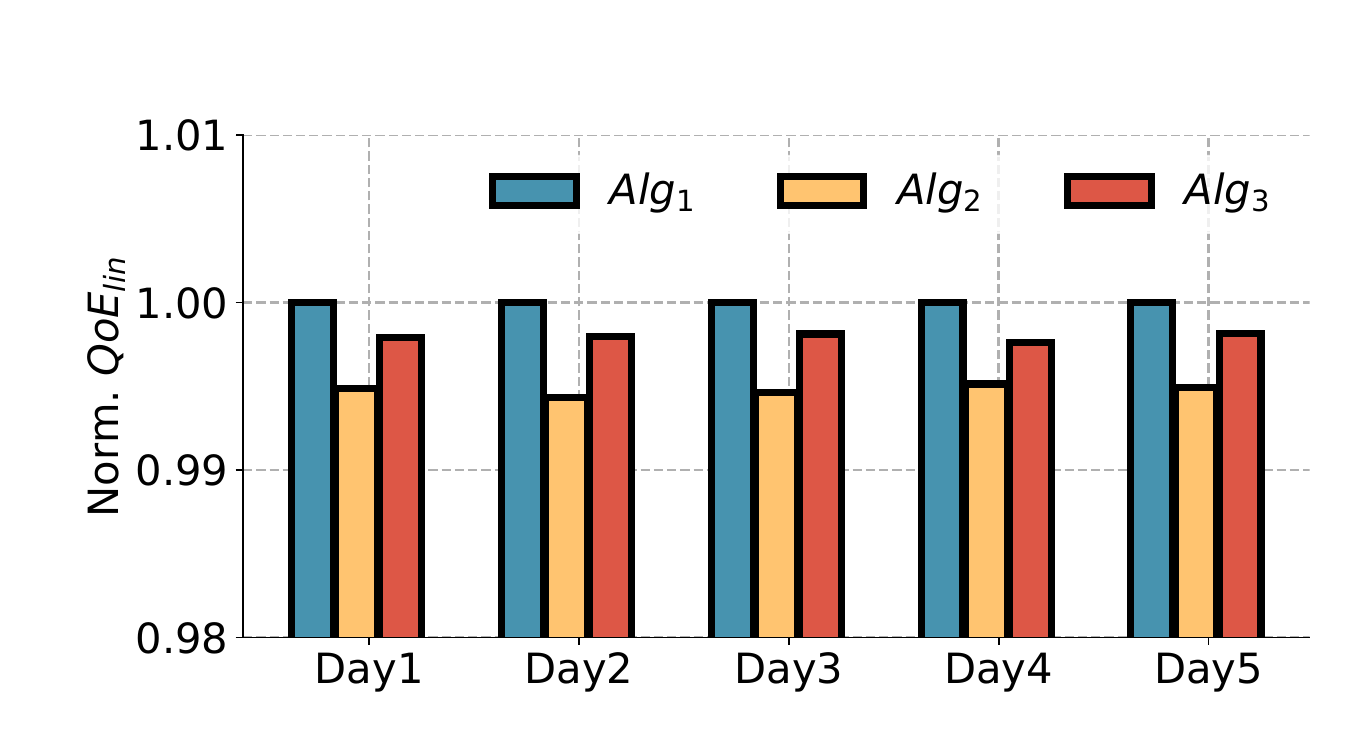}
    \label{img/intro/abnetnewqoe.pdf}
}
\subfigure[Overall Watch Time: Each alg achieved the highest watch time in different periods, showing no consistent superiority]{
    \includegraphics[width=0.45\linewidth]{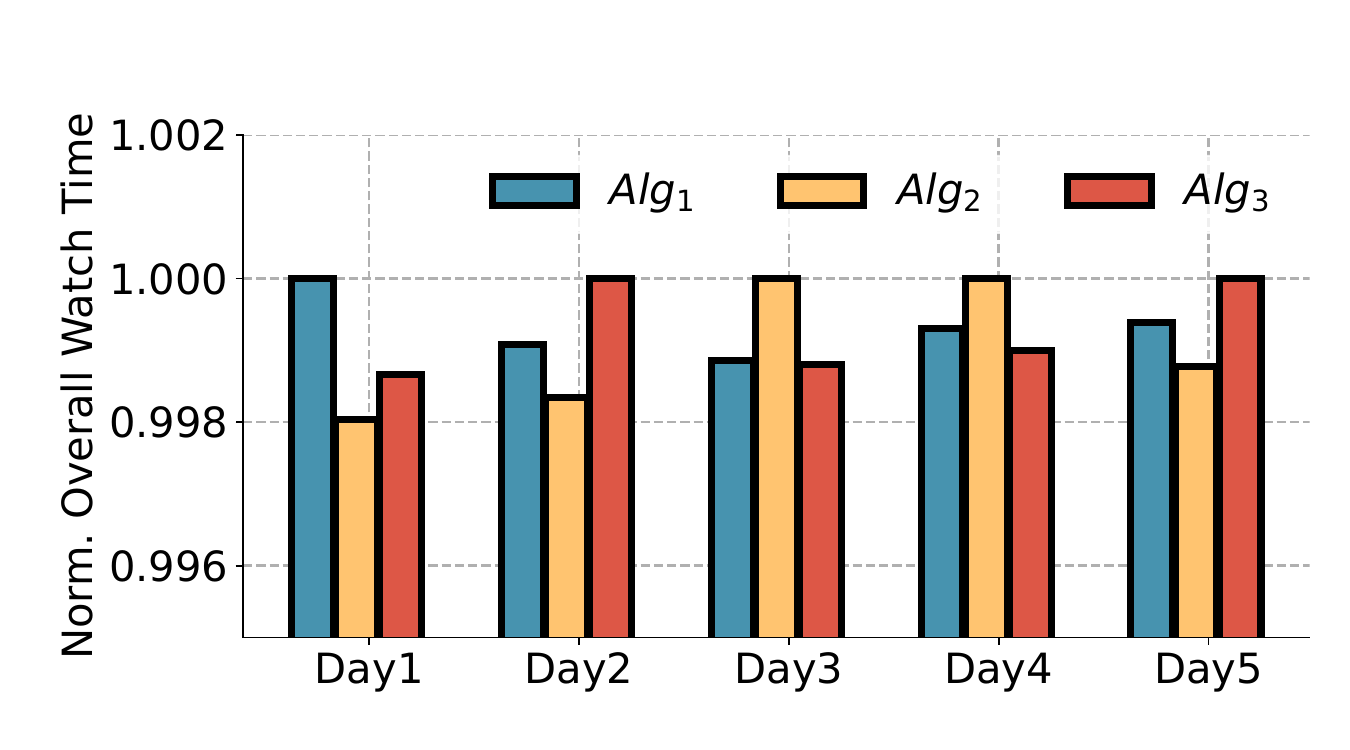}
    \label{img/intro/abnetnewt.pdf}
}
\vspace{-10pt}
        \caption{QoS, $QoE_{lin}$ and QoE obtained by algorithms with different optimization objectives in A/B test}
        \label{different optimization objectives in A/B test}
        \vspace{-5pt}
\end{figure}
\begin{figure}
    \centering
      \subfigure[Bandwidth CDF ]{
        \includegraphics[width=0.45\linewidth]{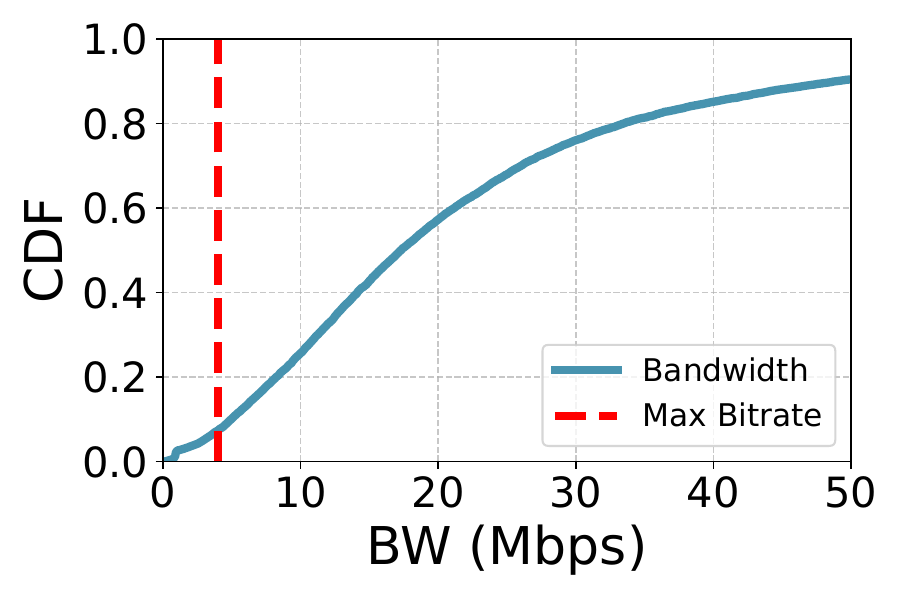}
        \label{img/intro/lingxi_cdf.pdf}}
        \subfigure[User Stall Counts in One Day]{
        \includegraphics[width=0.45\linewidth]{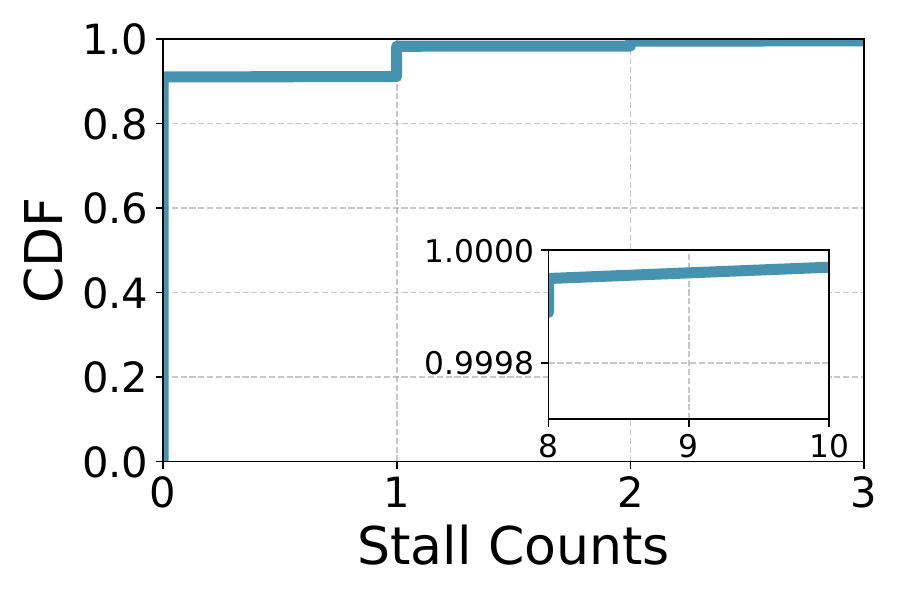}
        \label{img/intro/Stalls_CDF_with_zoom.pdf}
        }
                \vspace{-10pt}
        \caption{Optimization Opportunities in Production System}
        \label{fig: Room for Optimization in Production System}
        \vspace{-5pt}
\end{figure}

\section{From QoS to Personalized QoE}~\label{sec:QoS Impact on Exit Rate}

\subsection{When the garden is well-tended: QoS metrics meet their limits}~\label{When the garden is well-tended}
We conduct a comprehensive 5-day A/B testing experiment to evaluate algorithm performance across different optimization objectives, as shown in Figure~\ref{different optimization objectives in A/B test}. The experiment compares three variants: the production baseline algorithm and two modified versions with different optimization preferences—one prioritizing video quality ($Alg_3$) and another emphasizing stall time reduction ($Alg_1$).

The experimental results show that despite these algorithms having different optimization strategies, their differences on traditional QoS metrics are relatively limited: the relative improvement in video quality is restricted to approximately 0.5\%, while the maximum relative reduction in stall time is about 2\%. To better compare the performance differences among these algorithms, we adopt the widely-used linear QoE metric~\cite{yin2015control}:

\begin{equation}
QoE_{lin} = \sum_{k}q(Q_k) - \mu \sum_{k}T_k - \sum_{k}|q(Q_{k+1}) - q(Q_{k})|
\label{qoe}
\end{equation}

where $q(Q_k)$ represents the video quality when the $k$-th video segment selects bitrate level $Q_k$, $T_k$ represents the stall time of the $k$-th segment, and $\mu$ represents the weight parameter for stall penalty, which we set to the maximum video quality value.

The $QoE_{lin}$ calculation results indeed reflect the technical differences among algorithms, with $Alg_1$, which prioritizes stall time reduction, achieving the highest score. However, when we observe the real user engagement metric—total watch time, we find that all three algorithms achieve the highest watch time at different periods, showing no statistically significant differences.

This experimental study reveals two key findings. First, the overall QoS optimization space for existing algorithms is approaching saturation. Second, as modern adaptive streaming algorithms achieve high performance levels, traditional QoS metrics and $QoE_{lin}$ are becoming increasingly ineffective at distinguishing algorithm QoE performance. These findings suggest that strategies focused on optimizing system-wide QoS metrics may have approached their performance boundaries.

Analysis of production system online logs (presented in Figure~\ref{fig: Room for Optimization in Production System}) reveals the current state of streaming optimization. Leveraging advances in mobile network infrastructure~\cite{narayanan20215g,wifi6}, merely 10\% of users experience average bandwidth below maximum video bitrate requirements. Moreover, over 90\% of users achieve stall-free playback in their daily usage, while more than 99\% of users encounter no more than two stall events per day. 

The examination of A/B testing outcomes and production logs indicates that system-level QoS optimization approaches its performance ceiling. However, in large-scale streaming systems, low-bandwidth users comprising more than 10\% of the total user base still constitute a significant absolute population, indicating substantial optimization potential. This realization necessitates a paradigm shift toward individual-level QoE optimization as the next frontier.
\begin{figure}
    \centering
      \subfigure[Video Quality]{
        \includegraphics[width=0.45\linewidth]{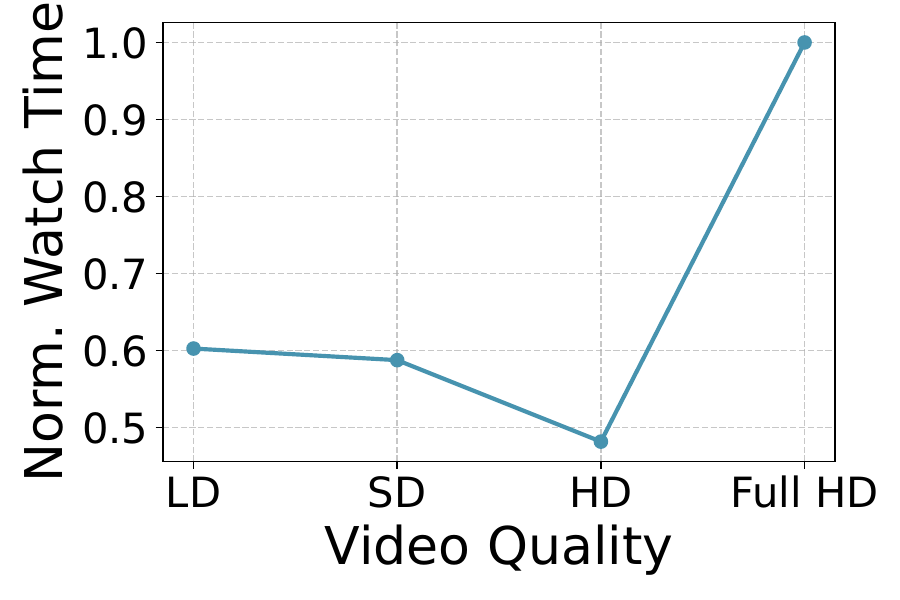}
        \label{img/mot/quality_analysis.pdf}
        }
        \subfigure[Stall Time]{
        \includegraphics[width=0.45\linewidth]{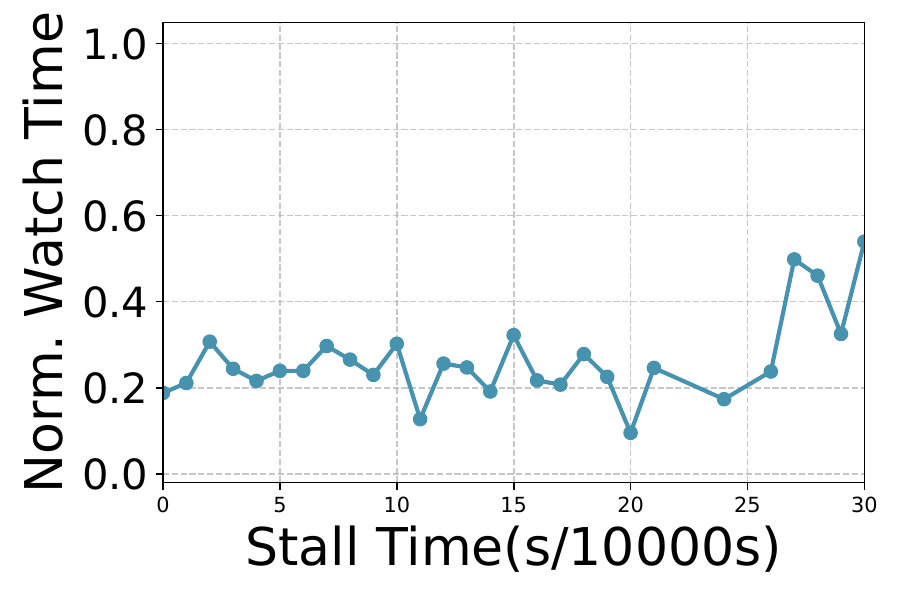}
        \label{img/mot/each_stall_time_analysis.pdf}
        }   
        \vspace{-10pt}
        \caption{The Impact of QoS Metrics on Watch Time}
        \label{fig: The Impact of QoS metrics on Watch Time}
        \vspace{-5pt}
\end{figure}

\subsection{Exit Rate: Fine-grained User-Level QoE}~\label{Exit Rate: QoE Metric for Analysi}
QoE, as a subjective metric, relies on users' evaluative feedback regarding their watch experiences across diverse video playback trajectories~\cite{itu1999subjective}. However, in production environments, the process of soliciting user ratings inherently affects the watch experience, rendering subjective evaluation methods impractical for implementation~\cite{zhang2022enabling}. Consequently, there is an increasing emphasis on leveraging objective metrics directly extractable from system logs, such as watch time and exit rate, which have emerged as pragmatic standards for QoE assessment~\cite{dobrian2011understanding}.

Our analysis encompasses 1.5 million playback trajectories from the production environment. Each trajectory represents a video playback session, including user IDs (with privacy data removed), watch timestamps, total video lengths, user watch time, and information regarding each video segment, such as buffer size, bitrate levels, segment sizes, download time, and stall time. 
While watch time represents the most intuitive and significant metric, its applicability is restricted to completed sessions or predetermined statistical intervals (e.g., daily aggregation). This limitation in data collection introduces substantial randomness when analyzing individual users' daily watch duration, potentially leading to conclusions that deviate from established cognitive patterns, as shown in Figure~\ref{fig: The Impact of QoS metrics on Watch Time}. Moreover, as a long-term metric, watch time exhibits limited sensitivity to short-term playback dynamics, explaining its predominant application in CDN optimization research~\cite{cheng2023rebuffering}. 

\begin{figure}
    \centering
      \subfigure[Video Quality]{
        \includegraphics[width=0.45\linewidth]{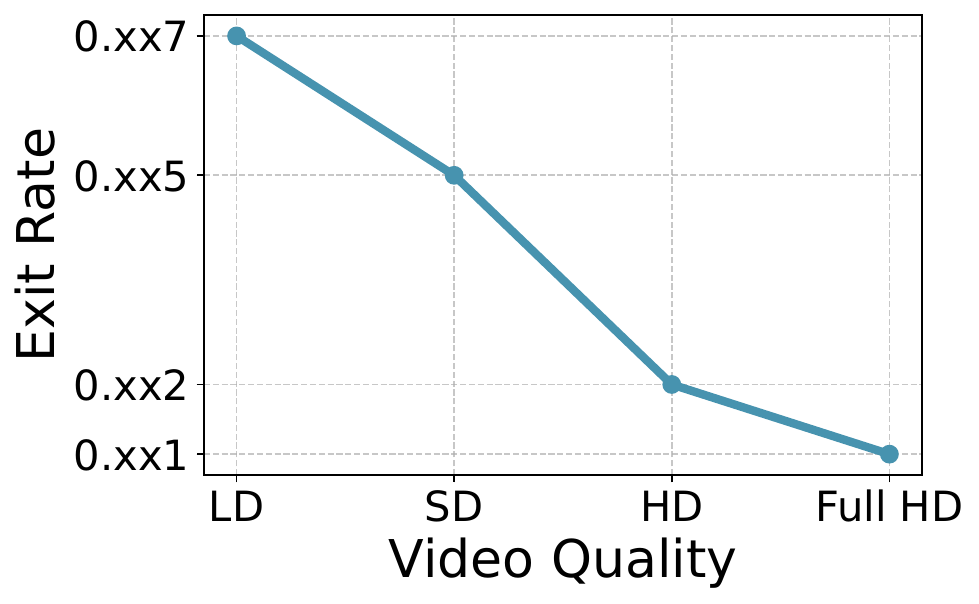}
        \label{img/mot/resolution_value_chart.pdf}
        }
        \subfigure[Video Smoothness]{
        \includegraphics[width=0.45\linewidth]{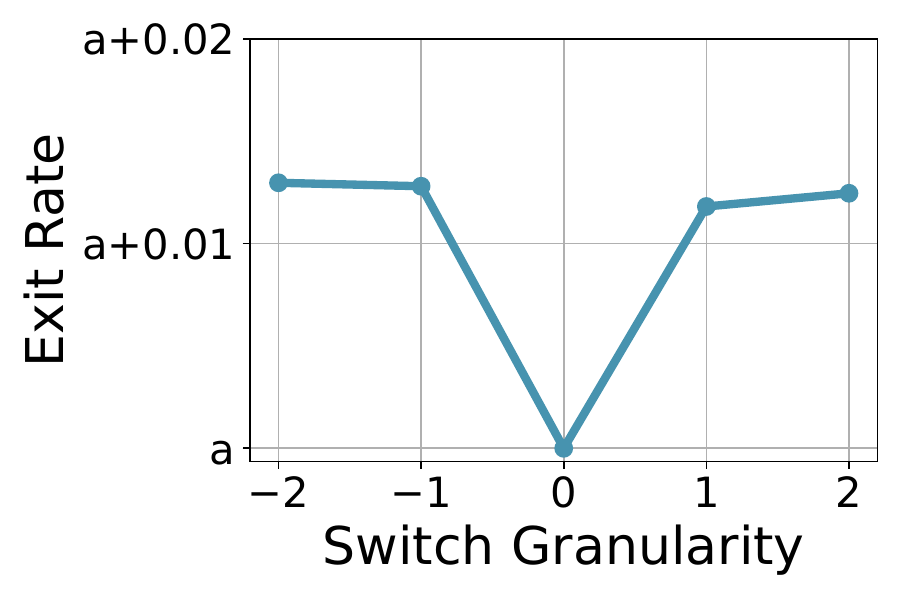}
        \label{img/mot/switch_1.pdf}
        }
     \subfigure[Stall Time Overall]{
        \includegraphics[width=0.45\linewidth]{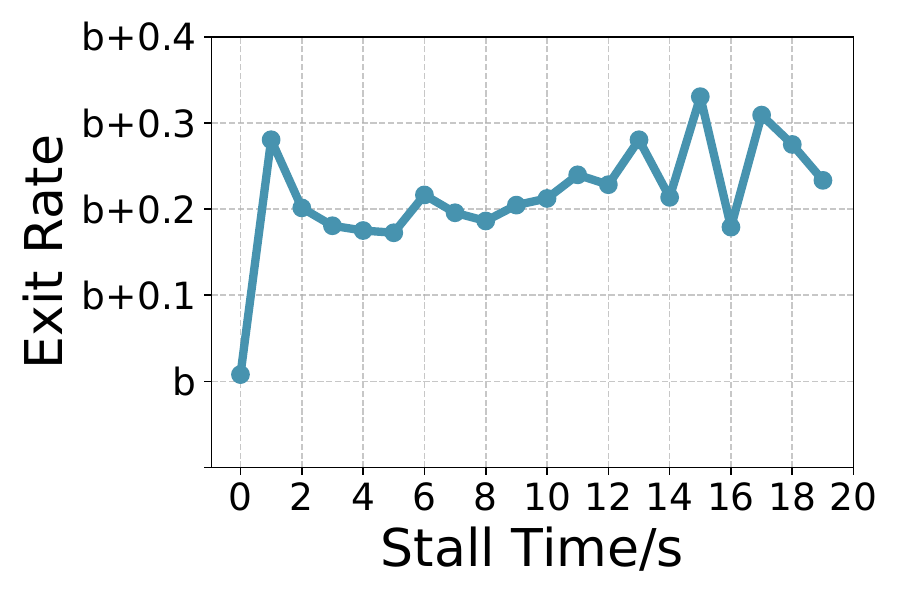}
        \label{img/mot/dataset_1.pdf}
        }
        \subfigure[Stall Time with Compound Effects]{
        \includegraphics[width=0.45\linewidth]{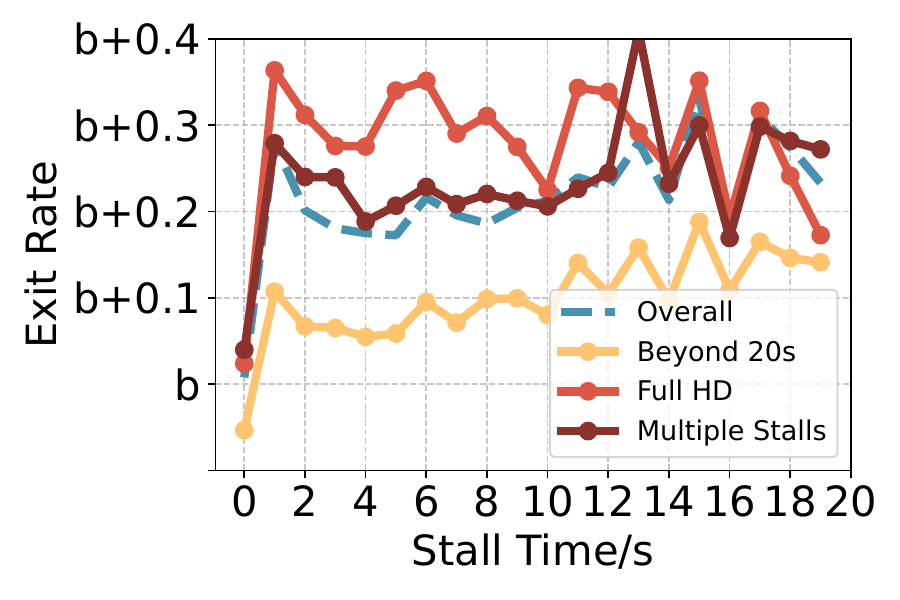}
        \label{img/mot/dataset_2.pdf}
        }
       
         \vspace{-10pt}
        \caption{The Impact of QoS metrics on Exit Rates}
        \label{fig: The Impact of QoS metrics on Exit Rates}
        \vspace{-5pt}
\end{figure}

In contrast, the user exit rate represents the probability of user exit at the individual video segment level, serving as another form of user engagement measure. This metric enables more efficient data accumulation for statistical analysis and demonstrates stronger responsiveness to short-term playback behavior variations. This characteristic makes it suitable for research in ABR algorithms. Given these advantages, we adopt the exit rate as our primary QoE metric.
Figure~\ref{fig: The Impact of QoS metrics on Exit Rates} illustrates the relationship between various QoS metrics and exit rates, focusing on three fundamental components of $QoE_{lin}$: video quality, video smoothness, and stall time.

\textbf{Video Quality}  
We analyze four quality tiers: Low Definition (LD), Standard Definition (SD), High Definition (HD), and Full HD, as shown in Figure~\ref{img/mot/resolution_value_chart.pdf}. The results confirm an inverse correlation between video quality and exit rates. The marginal effect on exit rates diminishes notably between HD and Full HD, indicating reduced user sensitivity at higher quality tiers. However, the absolute difference in exit rates across quality levels remains quite limited, reaching only $10^{-3}$ order of magnitude.

\textbf{Video Smoothness}  
Figure~\ref{img/mot/switch_1.pdf} depicts the relationship between video smoothness and exit rates. Using non-switching sessions as the baseline (denoted as "a"), we evaluate exit rates under various quality transition patterns. The analysis reveals that quality switches consistently correlate with increased exit rates. While quality degradation shows a slightly higher impact than quality enhancement, the most significant difference exists between sessions with and without transitions. The magnitude of smoothness-related effects on exit rates approaches $10^{-2}$.

\textbf{Stall Time}  
Figure~\ref{img/mot/dataset_1.pdf} demonstrates the relationship between stall time and exit rates (from 0s with baseline exit rate "b"). Stall events exhibit a substantial impact on exit rate, with a generally monotonic positive correlation. Since long-time stall events are infrequent in online environments, their impact on exit rates is prone to interference from individual user engagement, leading to some fluctuations in the curve. However, this overall upward trend remains largely unaffected. The influence of different stall time on exit rates typically falls within the $10^{-1}$ magnitude range, with the maximum differential reaching approximately 0.3.

\textbf{Compound Effects}  
Consistent with previous QoE research~\cite{duanmu2018sqoe3,sqoe4}, the relationship between QoS metrics and exit rates exhibits complex interdependencies as Figure~\ref{img/mot/dataset_2.pdf}. Using stall time as the primary variable, we investigate its compound interactions with watch time, video quality, and stall counts. Analysis of sessions exceeding 20 seconds reveals significantly reduced exit rates compared to overall data, suggesting enhanced stall tolerance with prolonged engagement. The correlation between stall time and exit rates in full HD scenarios indicates that users' tolerance for stall diminishes as video quality increases. Further analysis of multiple-stall scenarios indicates substantially higher exit probabilities compared to single-stall events. These analyses indicate that the relationship between QoS metrics and exit rates is intricate and may be challenging to accurately characterize with specific formulas.

\noindent \textbf{Takeaway 1:}
The constituent components of $QoE_{lin}$--video quality, video smoothness, and stall time--demonstrate hierarchical influence magnitudes on exit rates at $10^{-3}$, $10^{-2}$, and $10^{-1}$, respectively. Considering that user exits can be significantly influenced by irrelevant factors such as content, personalized modeling of QoS metrics with smaller impact values is likely to be overshadowed by these unrelated fluctuations. Consequently, we implement a hybrid modeling framework that combines personalized modeling for stall events with universal modeling for other metrics in exit rate prediction. The correlation between stall events and exit behavior depends on multiple factors and exhibits complex relationships, requiring strong capabilities in capturing intricate patterns and nonlinear interactions. Neural networks, with their inherent ability to model complex relationships and handle high-dimensional feature interactions~\cite{zhang2019deep}, emerge as an ideal tool for personalized stall response modeling.

\begin{figure}
  \centering
      \subfigure[CDF of average tolerable stall time and difference between Day 1 and Day 2]{
        \includegraphics[width=0.43\linewidth]{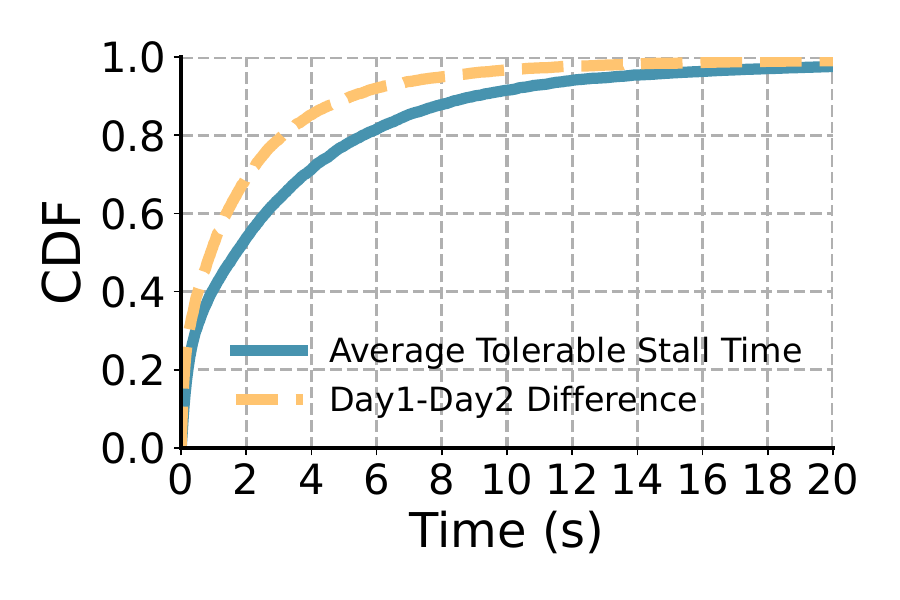}
        \label{CDF of Average Tolerable Stall Time}
        }
        \subfigure[User cases when encountering different stall times, categorized as sensitive, sensitive to threshold, and insensitive]{
        \includegraphics[width=0.43\linewidth]{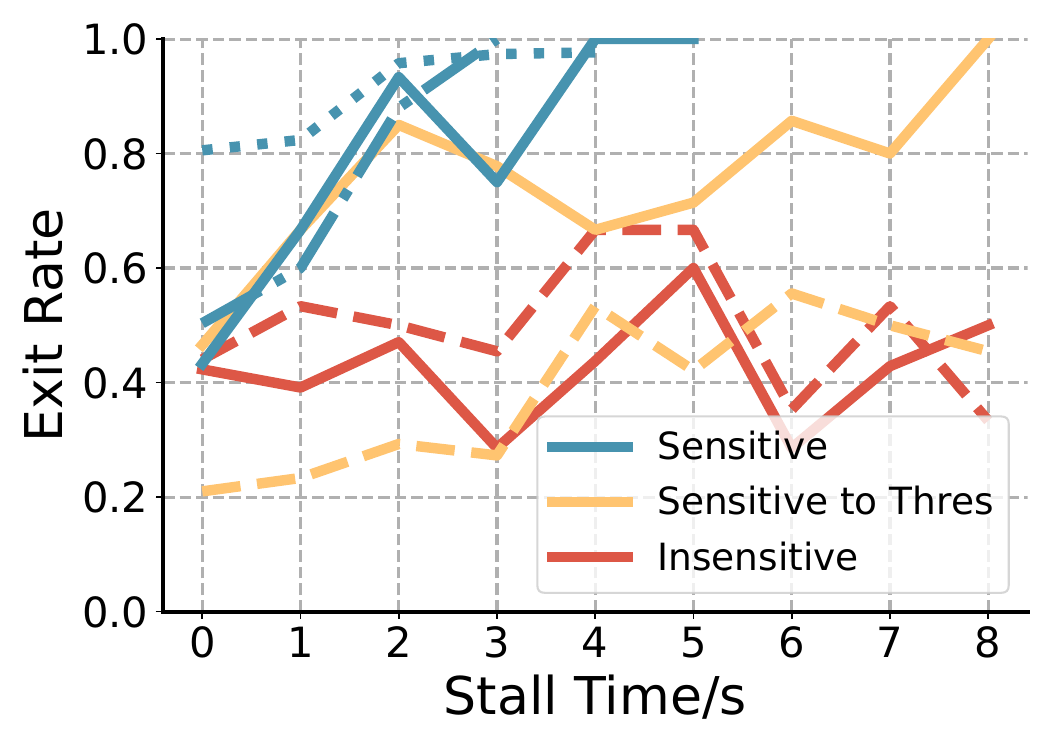}
        \label{Some User Case When Meets Different Stall Tim}
        }
                \vspace{-10pt}
         \caption{Personalized Perception of Stall}
        \label{Personalized Perception of Stall}
        \vspace{-5pt}
\end{figure}
\subsection{A Thousand Faces: Personalized Perception of Stall Time}~\label{sec:Online Log Analysis for QoS}

We conduct an in-depth analysis of the personalized relationship between stall time and exit rate among online users. In Figure~\ref{CDF of Average Tolerable Stall Time}, we present the Cumulative Distribution Function (CDF) of average tolerable stall time for active users who maintained continuous viewing sessions. The CDF analysis reveals heterogeneity in stall tolerance thresholds: approximately 20\% of users exhibit minimal tolerance to stall events, while another 20\% demonstrate resilience to stall exceeding 5 seconds, and about 10\% maintain engagement even with stall periods surpassing 10 seconds. Further analysis of day-to-day variations in tolerance time reveal a distinctive distribution pattern: most users maintain consistent tolerance levels with minimal fluctuations; about 20\% exhibit variations ranging from 2 to 4 seconds; and the rest demonstrate variations following a long-tail distribution. These findings indicate that user tolerance characteristics predominantly maintain stability while exhibiting certain temporal dynamics.

In Figure~\ref{Some User Case When Meets Different Stall Tim}, We present several users with watch times near the 90th percentile, illustrating the heterogeneity in exit rates when facing different stall times. Individual response patterns vary significantly: some users show rapidly increasing exit rates with longer stall times, others demonstrate threshold-based sensitivity, while certain users exhibit varied responses across different stall times.

\noindent \textbf{Takeaway 2:} The correlation between stall events and exit patterns exhibits significant individual heterogeneity, with most users maintaining notable stability. This phenomenon establishes the foundation for implementing user-level QoE optimization strategies. Furthermore, this correlation demonstrates dynamic evolutionary characteristics, indicating the necessity of adopting dynamic modeling approaches to capture these complex behavioral patterns.
\section{\texttt{LingXi} Design}\label{sec:sys}
\begin{figure}
    \centering
        \includegraphics[width=0.98\linewidth]{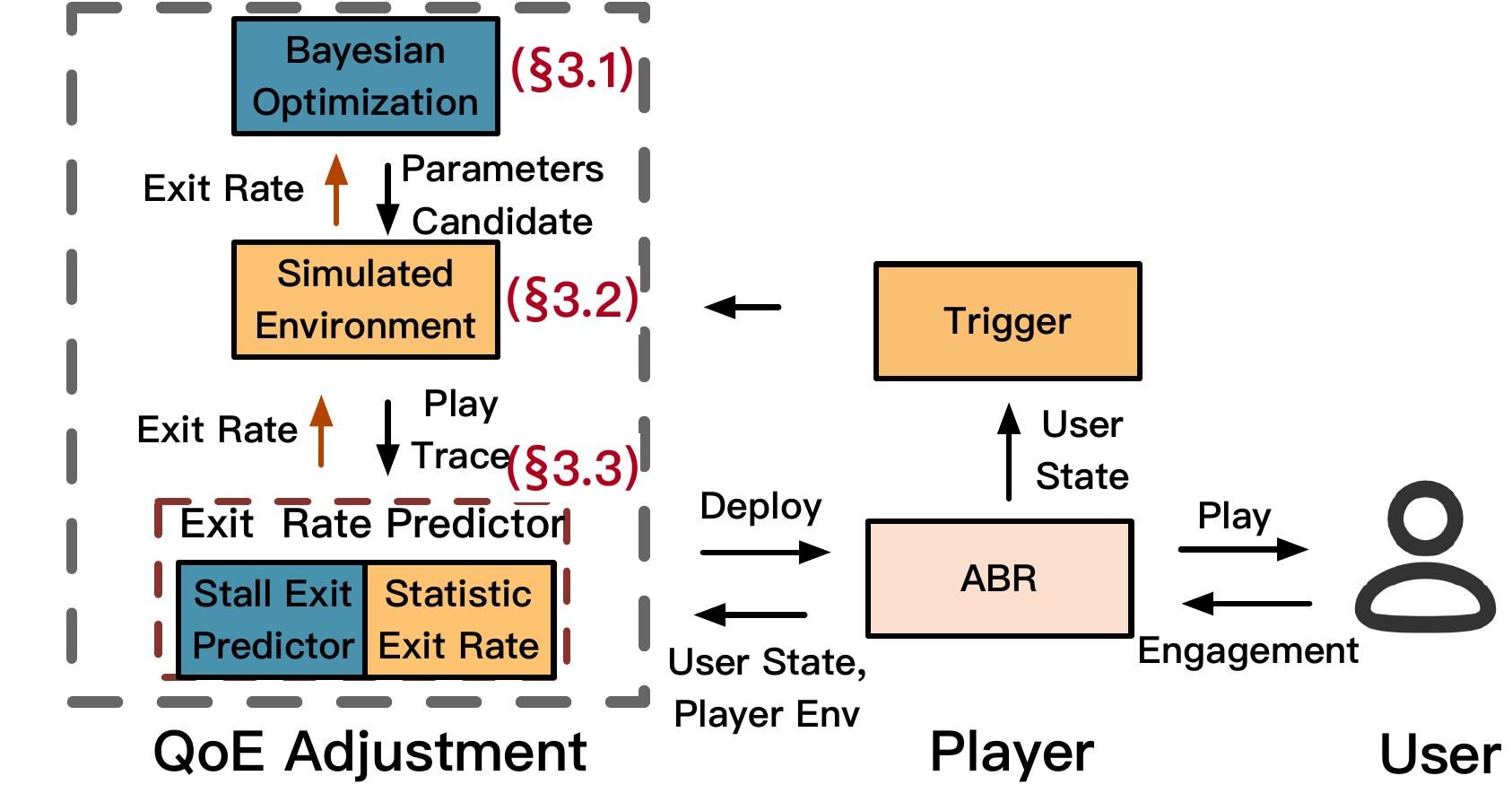} 
        \vspace{-5pt}
         \caption{\texttt{LingXi} System Overview}
         \vspace{-5pt}
         \label{LingXi system overview}
\end{figure}

\texttt{LingXi}'s architecture is illustrated in Figure~\ref{LingXi system overview}. \texttt{LingXi} leverages production environment data, combining playback logs and user engagement metrics to facilitate dynamic optimization of ABR algorithm objectives. We first introduce online Bayesian optimization methods in Sec~\ref{sec:Bayesian Optimization} to explore and determine ABR candidate parameters for QoE optimization. Sec~\ref{sec:Monte Carlo Alignment Perception} discusses how Monte Carlo sampling methods align the impact of bitrate selection on both immediate and long-term user exit behaviors, generating playback traces based on determined parameters. Finally, Sec~\ref{sec:Exit Rate Predictor} introduces an exit rate predictor based on hybrid modeling of personalized neural networks and overall statistical data, quantifying user experience under different parameter settings through sampled traces.

\subsection{Online Bayesian Optimization}~\label{sec:Bayesian Optimization}
Since \texttt{LingXi} is based on users' historical playback processes and engagement, obtaining the optimal QoE parameters for online matching through offline search is not feasible. Additionally, the relationship between QoE parameters and exit rates cannot be represented in an exact analytical form. Therefore, we treat \texttt{LingXi} as an online black-box optimization problem, where the inputs are QoE parameters \( x \), bandwidth distribution \( N \), current user state~\( S \), and player environment ~\( E_{player} \). Our objective is to determine an optimal \( x^* \) that minimizes the overall exit rate objective function \( f_{R_{exit}} \) over the domain \( X \) as:
\begin{equation}
x^* \in \text{argmin}_{x \in X} f_{R_{exit}}(x, N,S, E_{player})
\end{equation}

Given the production system's constraints on convergence speed and computational overhead, we choose Online Bayesian Optimization (OBO)~\cite{williams2006gaussian} as our optimization algorithm. In contrast to expert-defined static parameters, OBO facilitates more sophisticated modeling of user-specific QoS sensitivity patterns. OBO uses historical trial data to fit a surrogate model (e.g., Gaussian distribution) that approximates the objective function. By maximizing an acquisition function (e.g., rate of improvement), we determine the next QoE parameter \( x \) to explore and iteratively optimize the objective function.

To accommodate heterogeneous user sensitivities to QoS,  we apply OBO independently for each user. The optimization process initializes with default parameters and, upon activation of the QoE adjustment mechanism, leverages previously optimized configurations as initialization points for subsequent iterations. This adaptive strategy ensures continuous optimization while maintaining responsiveness to temporal variations in network conditions and user engagement patterns. The brief implementation of OBO within the \texttt{LingXi} framework is delineated in Algorithm~\ref{alg:obo}

\begin{algorithm}
    \SetAlgoLined
    \caption{Online Bayesian Optimization in \texttt{LingXi}}
    \label{alg:obo}
    
    \KwIn{Default QoE parameters $x_d$, max sampling time $T_s$, triggering threshold $\eta$, bandwidth model $N$, user states $S$, player environment $E_{player}$}
    \KwOut{Optimized QoE parameters $x^*$}

    ABR.init($x_d$)\;
    $x^* \leftarrow x_d$\;
    $R_{min\_exit} \leftarrow \infty$\;
    
    \While{streaming\_session\_active}{
        Run ABR algorithm, $N$.update(), $S$.update(), $E_{player}$.update()\;
        
        \If{stall\_count > $\eta$}{
            \tealcomment{Trigger OBO Process}\\
            OBO.init($x^*, N, S, E_{player}$)\;
            
            \While{Sample Time $< T_s$}{
                $x \leftarrow$ OBO.next\_candidate()\;
                $R_{exit} \leftarrow$ EvaluateParameters($x, N, S, E_{player}$)\;
                \tealcomment{Detailed in Algorithm \ref{alg:mc_simulation}}
                
                OBO.update($x, R_{exit}$)\;
                
                \If{$R_{exit} < R_{min\_exit}$}{
                    $R_{min\_exit} \leftarrow R_{exit}$\;
                    $x^* \leftarrow x$\;
                }
            }
            ABR.update($x^*$)\;
            $R_{min\_exit} \leftarrow \infty$\;
           
        }
    }
\end{algorithm}

\subsection{Monte Carlo Alignment between Immediate and Long-term Impacts}\label{sec:Monte Carlo Alignment Perception}

The current bitrate selection not only affects the immediate QoE but also influences the subsequent playback process, thereby impacting future long-term QoE. Therefore, our optimization framework must consider both the immediate exit rate and its future implications. Given the uncertainty in network conditions and user behaviors, the Monte Carlo simulation presents an ideal approach for handling such stochastic processes and exploring various possible future scenarios efficiently. Therefore, we employ Monte Carlo sampling to simulate real user watch patterns and playback processes to align the immediate and future impacts on exit rates. We model the past bandwidth $C_{past}$ perceived by the client when downloading video segments as a normal distribution and sample from it to represent future bandwidth. Starting from the current player environment $E_{player}$, we utilize the candidate QoE parameters to adjust ABR and conduct virtual playback through Monte Carlo sampling. The exit rate predictor computes segment-level exit rates based on user states. The parameters yielding minimal exit rates are subsequently deployed in online ABR operations. We present the complete Monte Carlo workflow in Algorithm~\ref{alg:mc_simulation}.

\textbf{Aligning with Online User Watch Patterns}  
Our sampling methodology aims to precisely approximate real-world user watch patterns. We run $M$ samples with maximum duration $T_{sample}$ per sample, where $T_{sample}$ is set to the average length of online videos and the total simulation time $M \times T_{sample}$ aligns with the mean watch time of daily active users during peak periods. User exit behavior is modeled by calculating segment-level exit rates using user states that include long-term user engagement and playback information, with exit decisions made randomly based on these probabilities. This systematic approach enables us to replicate authentic viewing patterns and capture the occurrence probabilities of critical events, including stalls and bitrate transitions. Through this framework, we can establish correlations between immediate exit rates and their subsequent impact on future exit rates. 

\textbf{Aligning with Online Playback Processes}  
We reference previous classic works~\cite{yin2015control} and production environment settings to model the playback process. This model serves as the player environment transition function, with the core simulation rules that dictate how the player's environment evolves from one segment to the next formally defined in Equation~\ref{equ:simulator}:

\begin{equation}
\begin{aligned}
& C_k \sim \mathcal{N}(\mu_{C_{past}}, \sigma_{C_{past}}^2) \\
& B_{k+1} = \left(\left(B_k - \frac{d_k(Q_k)}{C_k}\right)_+ + L - \delta t_k\right)_+, \\
& \delta t_k = \max(B_{k+1} - B_{\max}, 0) + \text{RTT} \\
& B_k \in [0, B_{max}], \quad B_{max} = f(\mathcal{N}(\mu_{C_{past}}, \sigma_{C_{past}}^2)) \\
& Q_k \in \mathcal{Q}, \quad \forall k = 1,\cdots,K. 
\end{aligned}
\label{equ:simulator}
\end{equation}

As shown in the equation, at each virtual time $t_k$, we first sample the current bandwidth $C_k$ based on the bandwidth model. We then calculate the download time for the selected bitrate $Q_k$, where the video segment size is $d_k(Q_k)$, using the formula $\frac{d_k(Q_k)}{C_k}$. Next, we update the buffer $B_{k+1}$ based on download time, waiting time $\delta t_k$, and video segment length $L$. The waiting time is calculated as the portion of $B_{k+1}$ that exceeds $B_{max}$, plus the round-trip time (RTT). The online adjustment of $B_{max}$ is a function related to bandwidth. Through this model, each sample starts from the initial state and continues until the playback time reaches $T_{sample}$ or an exit event occurs. The final exit rate is calculated as the ratio of total exit events to total segments watched across all samples: $R_{exit} = \frac{\text{exited\_count}}{\text{watched\_count}}$.

\begin{algorithm}
    \SetAlgoLined
    \caption{Implementation of \texttt{EvaluateParameters} via Monte Carlo Simulation}
    \label{alg:mc_simulation}

    \KwIn{
        Candidate QoE parameters $x$, 
        Bandwidth model $N$,
        Current user state $S$, 
        current player environment $E_{player}$,
        Number of Monte Carlo samples $M$, 
        Max duration per sample $T_{sample}$
    }
    \KwOut{Estimated exit rate $R_{exit}$ for parameters $x$}

    $\text{exited\_count} \leftarrow 0$, $\text{watched\_count} \leftarrow 0$\;
    
    \For{$m \leftarrow 1$ \KwTo $M$}{
        $S_{sim} \leftarrow S$, $E_{sim} \leftarrow E_{player}$, $t_{sim} \leftarrow 0$\;
        $\text{has\_exited} \leftarrow \text{false}$\;
        \While{$t_{sim} < T_{sample}$ \textbf{and not} 
        $\text{has\_exited}$}{
         \tealcomment{Terminates upon reaching the sampling time or when a random exit event occurs}
            $p_{exit} \leftarrow \text{ExitPredictor.predict}(S_{sim})$\;
            \tealcomment{Predict instantaneous exit probability}

            \If{$\text{random}(0, 1) < p_{exit}$}{
                $\text{exited\_count} \leftarrow \text{exited\_count} + 1$\;
                $\text{has\_exited} \leftarrow \text{true}$\;
                \tealcomment{The simulated user exits the video}
            }
           
            $C_k \leftarrow \text{Sample from } N(\mu_{C_{past}}, \sigma_{C_{past}}^2)$\;
            
            $Q_k \leftarrow \text{ABR.select\_bitrate}(S_{sim}, x)$\;
            \tealcomment{Apply the candidate QoE parameters to ABR}
            
            $E_{sim}.update(C_k, Q_k)$\;
            \tealcomment{Update player environment based on Equation \ref{equ:simulator}}
            
            $t_{sim} \leftarrow t_{sim} + L$\\
            $\text{watched\_count} \leftarrow \text{watched\_count} + 1$\;
            $S_{sim}.update()$\;
              \tealcomment{Update both short-term and long-term state}
        }
    }

    $R_{exit} \leftarrow \frac{\text{exited\_count}}{\text{watched\_count}}$\;
    \Return{$R_{exit}$}\;
\end{algorithm}

\subsection{Exit Rate Predictor}~\label{sec:Exit Rate Predictor}
\begin{figure}
    \centering
        \includegraphics[width=0.95\linewidth]{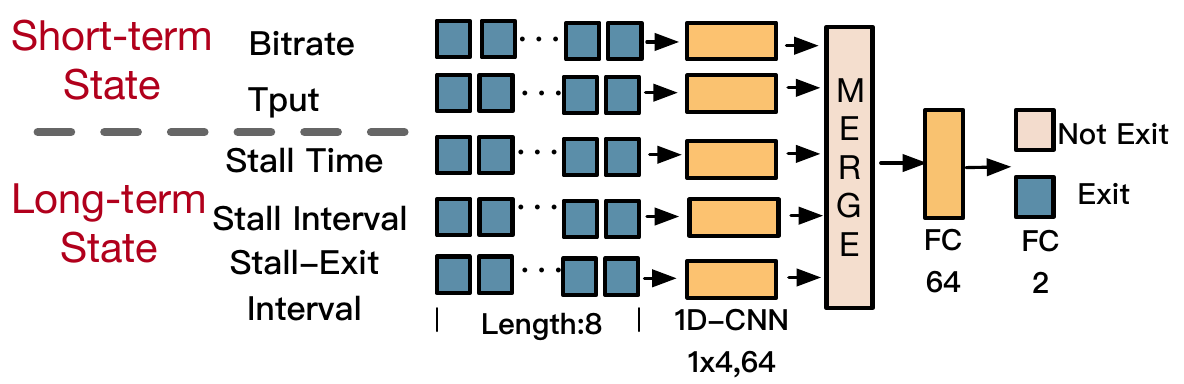} 
        \vspace{-5pt}
         \caption{Neural Network Architecture}
        \vspace{-5pt}
         \label{NN}
\end{figure}

Based on previous analyses, stall events emerge as the dominant determinant of exit rate. Consequently, we implement a hybrid modeling approach that combines personalized stall modeling with overall statistical (OS) modeling for video quality and smoothness metrics. Given the complex interdependencies between stall and other QoS parameters, we employ neural networks to capture these non-linear relationships as Figure~\ref{NN}. The exit rate $R_{exit}$ is thus formulated as:

\begin{equation}
\begin{split}
R_{exit} = 
\begin{cases}
NN(Stall) + OS(Quality,Smoothness), & \text{if stall } \\[8pt]
OS(Quality,Smoothness), & \text{else}
\end{cases}
\label{eq:c}
\end{split}
\end{equation}

\textbf{Dataset and Preprocessing}  
The dataset comprises approximately 100,000 entries, extracted from online logs exhibiting stall events. The dataset is partitioned into training and testing sets following an 80:20 stratification ratio. Given the class imbalance in the dataset, where the ratio of the continued watch to exits approaches 4:1 even in stall scenarios, we implement balanced sampling during model training. The methodology involves binary classification of the training set based on user engagement, followed by random undersampling of the majority class (continued watch) to achieve parity with the minority class (exits), thereby mitigating class distribution skewness.

\textbf{Input}  
To capture the complex relationship between stall and exit rates, we model this relationship across five dimensions: bitrate, throughput, past stall time, last stall interval, and last stall-exit interval. Bitrate reflects video quality and smoothness, while throughput represents overall network conditions. We record historical stall information by saving the stall time and the intervals since previous stalls and stall-triggered exits to capture user engagement. To balance the amount of information with computational resources, we set the matrix length to 8. The first two dimensions correspond to the last eight video segments, while the last three dimensions relate to stall events and user engagement. Through this method, we retain both short-term video playback information and long-term user engagement information within a single matrix.

\textbf{Output}  
We output a two-dimensional vector representing the probabilities of continuing to watch the video and exiting.

\textbf{Architecture}  
We utilize five 1D-CNN layers with 64 channels to extract state features, merging these vectors before inputting them into a 64-dimensional fully connected layer and a subsequent 2-dimensional fully connected layer. Finally, the neural network outputs a vector representing the exit rate through a softmax activation function. 

\textbf{Loss Function}  
In our approach, we model user engagement as a one-dimensional one-hot encoded vector, denoted as $R_{label}$. This vector is compared against the current model's output (the exit rate vector $R_{pred}$). To quantify the difference between these two vectors, we employ the cross-entropy function to calculate the loss. We formally defined as follows:

\begin{equation}
Loss = - \sum_{i=1}^{N} R_{label,i} \log(R_{pred,i})
\end{equation}

\section{Deployment}~\label{sec:Deployment}
\texttt{LingXi} demonstrates comprehensive compatibility with existing ABR systems. We implement the system through C++ integration into the production environment and detail its deployment through three fundamental aspects: integration methodology, activation criteria, and optimization mechanisms.

\textbf{Seamless Integration} \texttt{LingXi} implements a dual-layer state management strategy, utilizing a persistence mechanism to handle long-term and short-term states differently. For long-term states, the system serializes historical behavior-related data into HDF5 format files when the application terminates, while short-term states are initialized upon each startup. To minimize the impact on user experience, the restoration of long-term states occurs asynchronously after the initial screen rendering, thereby preventing I/O operations from affecting startup latency.
\texttt{LingXi} employs a priority-based dual-thread execution model where parameter adjustment processes run parallel to regular playback logic. The parameter adjustment tasks are allocated to low-priority background threads. Through thread priority scheduling and resource control, the system ensures that background computations do not interfere with the main playback process.

\begin{figure}
  \centering
      \subfigure[CDF of Daily Stall Counts]{
        \includegraphics[width=0.43\linewidth]{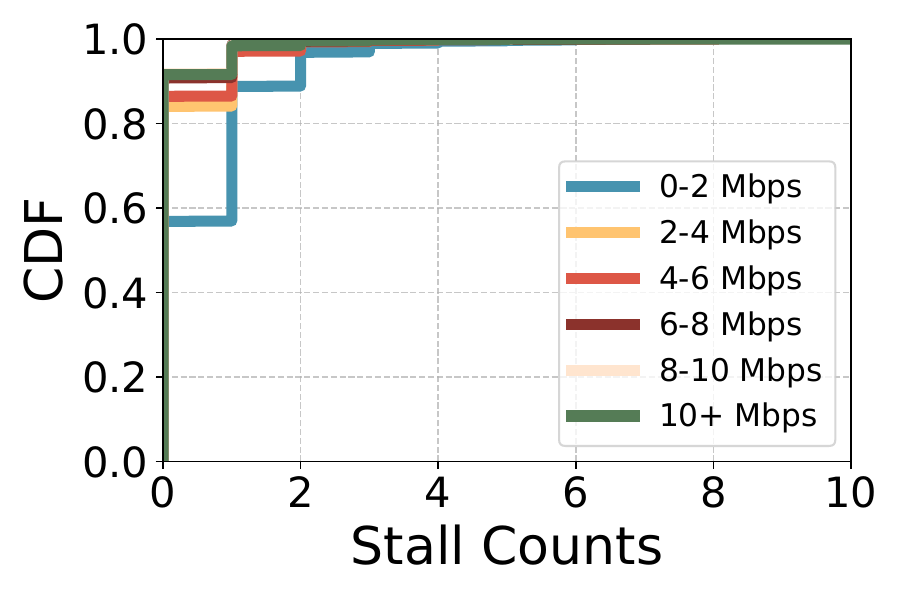}
        \label{CDF of Stall Counts}
        }
        \subfigure[Relationship between Stall Counts and Recall]{
        \includegraphics[width=0.43\linewidth]{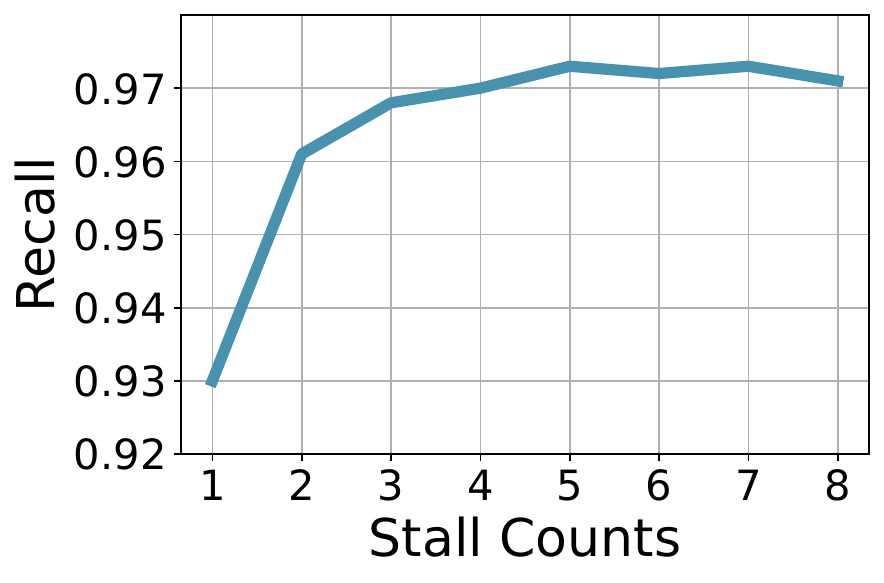}
        \label{Relationship between Stall Counts and Accuracy}
        }
                \vspace{-5pt}
         \caption{Trade-offs between Stall Counts and Recall}
        \label{Trade-offs between Stall Counts and Accuracy}
       \vspace{-5pt}
\end{figure}

\textbf{Trigger} The accumulation of historical user data enhances our capacity for user profile modeling and stall sensitivity analysis. However, the relatively low frequency of online stall events presents a fundamental trade-off between model accuracy and user coverage.
Figure~\ref{CDF of Stall Counts} illustrates the CDF of daily stall event frequency across different bandwidth segments. In high-bandwidth scenarios (>4000 kbps), over 95\% of users experience no stall events. Even in low-bandwidth conditions, the incidence of multiple stall events (>2) remains below 2\%, demonstrating the relative rarity of stall occurrences. Figure~\ref{Relationship between Stall Counts and Accuracy} depicts the correlation between stall event counts and model recall performance. While recall exhibits a general positive correlation with accumulated stall events, a notable discontinuity occurs between one and two stall events, showing approximately 3\% improvement. Although setting a higher threshold (e.g., eight stall events) could potentially enhance model accuracy, such a threshold would require extended data collection periods (weeks or months) for activation. Given the dynamic nature of QoS perception and the impracticality of prolonged data accumulation, we implement a threshold of two stall events as an optimal compromise between model performance and temporal responsiveness.

\textbf{Pruning}
To reduce overhead in the online system, we implement a dual-stage pruning strategy. The pruning mechanism operates at both the virtual playback level and the pre-playback assessment stage. During virtual playback simulation, we employ an early termination criterion: if the predicted exit rate under current parameters exceeds the minimum exit rate observed across alternative parameter sets, further evaluation of the current parameter configuration becomes redundant. At the pre-playback stage, we evaluate the probability of stall events based on bandwidth distribution characteristics. When the bandwidth distribution significantly exceeds maximum bitrate requirements, specifically when $\mu_{C_{past}} - 3\sigma_{C_{past}} > Q_{max}$, the likelihood of stall events becomes negligible. In such scenarios, where personalized optimization offers minimal potential benefit, the evaluation process is pruned.

\begin{figure}
    \centering
      \subfigure[Different Datasets]{
        \includegraphics[width=0.96\linewidth]{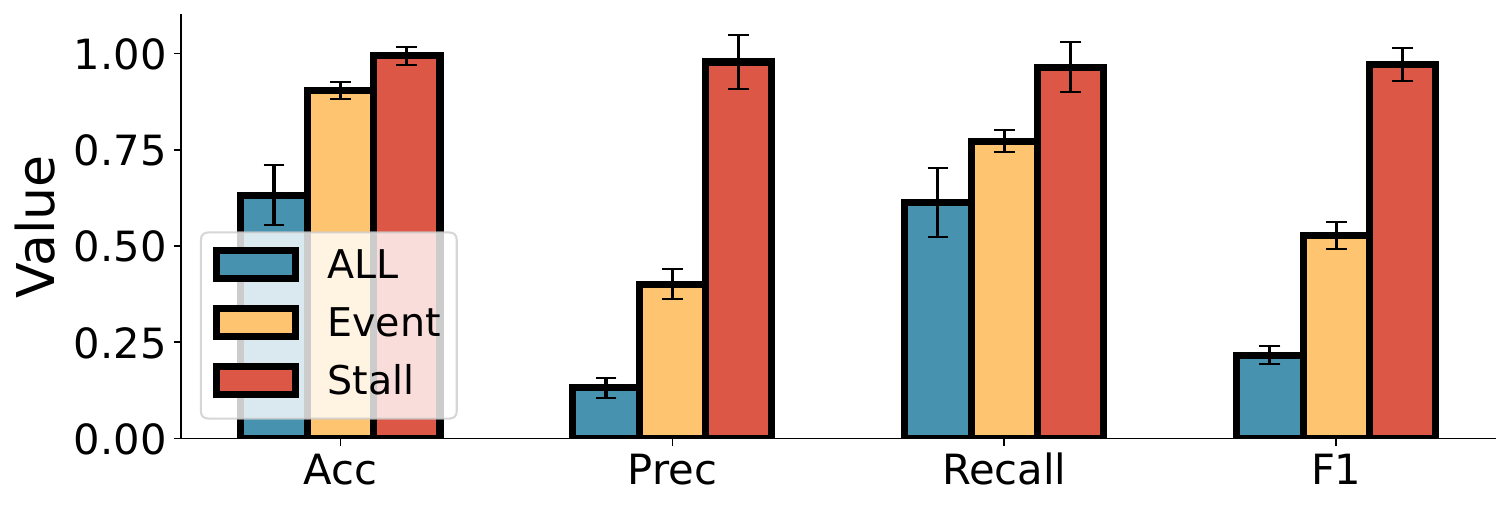}
        \label{img/eva/comparison1_with_f1.pdf}
        }
        \subfigure[W/WO Balanced Sampling]{
        \includegraphics[width=0.96\linewidth]{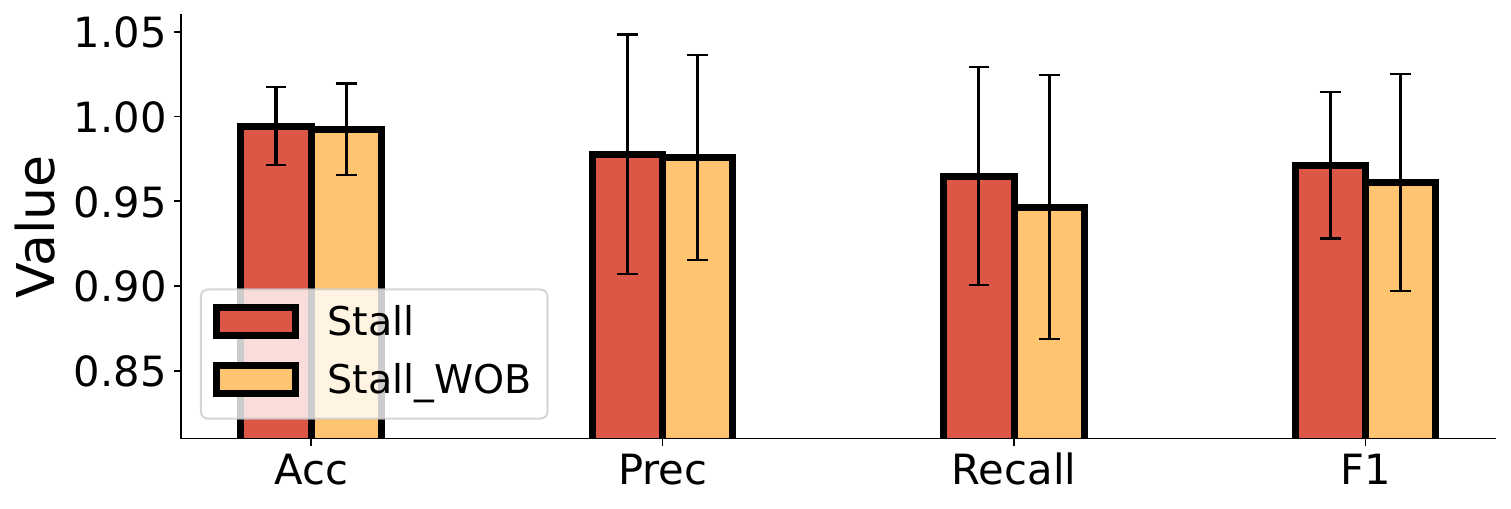}
        \label{img/eva/comparison2_with_f1.pdf}
        }
   \vspace{-5pt}
                \caption{Exit Rate Predictor in Different Setting}
        \label{Exit Rate Predictor in Different Setting}
       \vspace{-5pt}
\end{figure}

\begin{figure*}
    \centering
      \subfigure[Rule-based modeling with RobustMPC]{
        \includegraphics[width=0.3\linewidth]{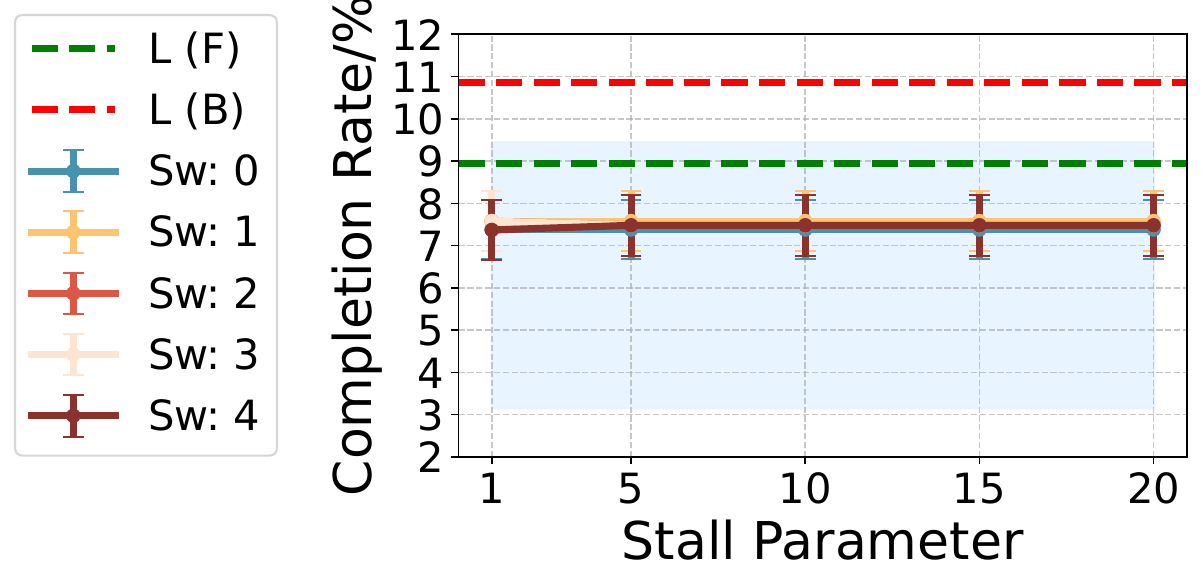}
        \label{Fixed User, MPC}
        }
        \subfigure[Rule-based modeling with Pensieve]{
        \includegraphics[width=0.215\linewidth]{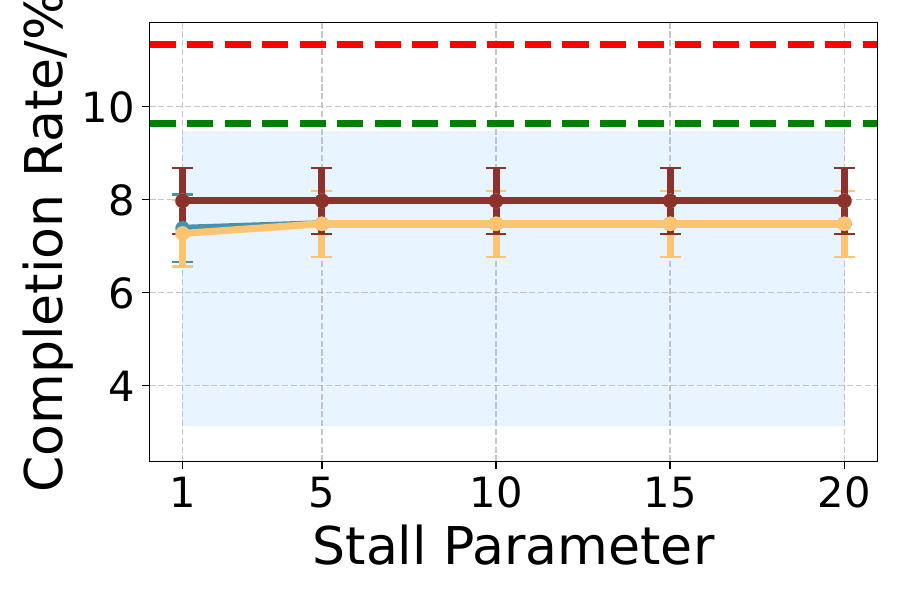}
        \label{Fixed User, Pensieve}
        }
     \subfigure[Data-driven modeling with RobustMPC]{
        \includegraphics[width=0.215\linewidth]{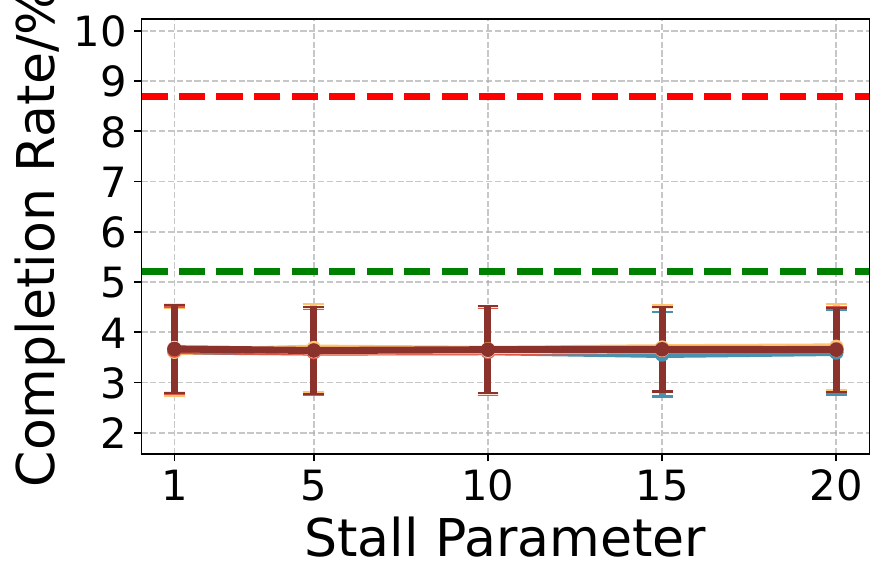}
        \label{User Model, MPC}
        }
        \subfigure[Data-driven modeling with Pensieve]{
        \includegraphics[width=0.215\linewidth]{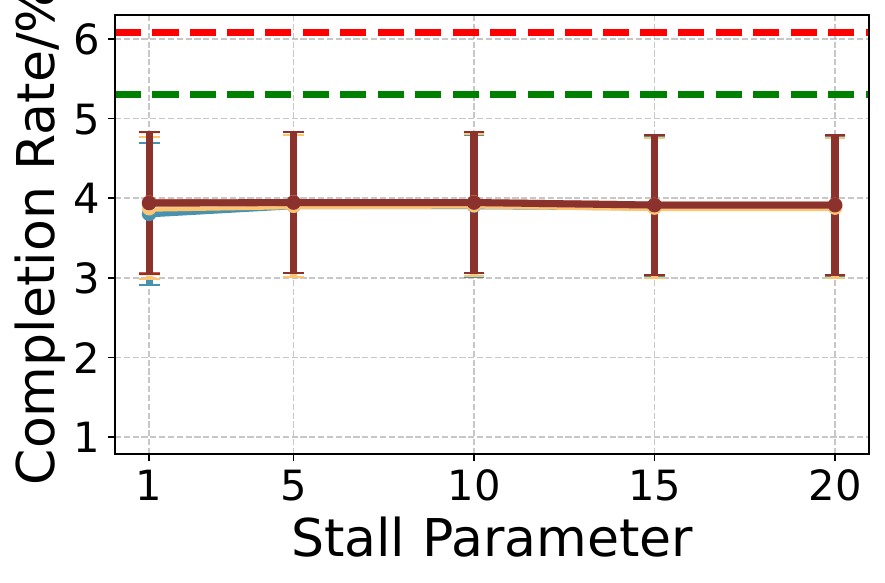}
        \label{User Model, Pensieve}
        }
          \vspace{-5pt}
                \caption{The Simulation Experiment of \texttt{LingXi}}
        \label{The Simulation Experiment of LingXi}
       \vspace{-5pt}
\end{figure*}
\begin{figure}
    \centering
      \subfigure[RobustMPC]{
        \includegraphics[width=0.43\linewidth]{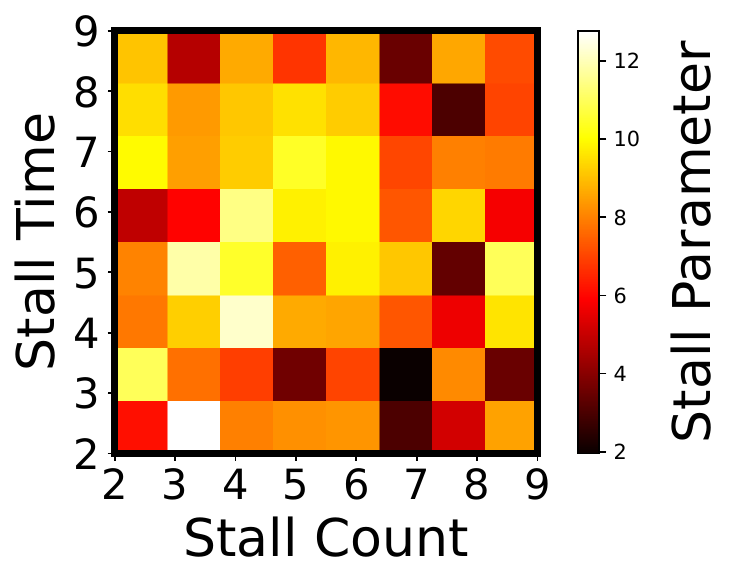}
        \label{img/eva/hot4.pdf}
        }
        \subfigure[Pensieve]{
        \includegraphics[width=0.43\linewidth]{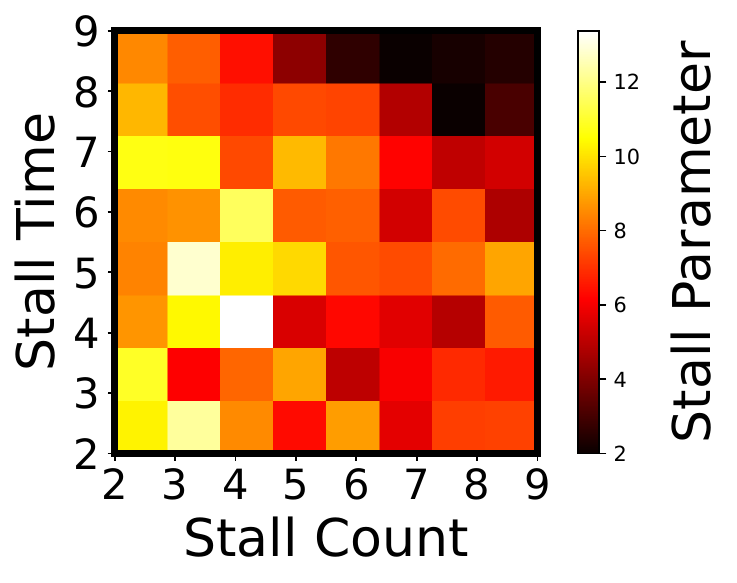}
        \label{img/eva/hot5.pdf}
        }
      \vspace{-5pt}
        \caption{Heatmap of Stall Parameters under Different Sensitivities}
        \label{Heatmap of Stall Parameters under Different Sensitivities}
        \vspace{-5pt}
\end{figure}
\section{Evaluation}~\label{sec:Evaluation}
First, we analyze the prediction accuracy of the exit rate and provide a reason for personalizing users' stall perception as detailed in Sec~\ref{sec:exit rate}. Subsequently, we conduct pre-deployment simulation experiments to validate the algorithm's effectiveness, as presented in Sec~\ref{sec:Simulation Evaluation}.  We model user preferences as two types: fixed preference rules and data-driven preference models. Next, we conduct a 10-day difference-in-differences A/B test in the production environment to verify the dual improvement in QoE and QoS as Sec~\ref{sec:A/B Evaluation}. Then, we analyze improvements for long-tail online users in low-bandwidth scenarios in Section~\ref{sec:long tail}. Finally, we perform a fine-grained analysis of online user engagement as~\ref{sec:Online User Study}.
\subsection{Exit Rate Prediction and Analysis}~\label{sec:exit rate}

\textbf{Different Dataset Composition}  
Figure~\ref{img/eva/comparison1_with_f1.pdf} illustrates the prediction results of predictors trained on different datasets. We employ identical dataset partitioning and sampling methods (80:20 ratio with balanced sampling). Each model undergoes five training iterations using different random seeds, with standard errors plotted as error bars. The results indicate that when using all states to predict exit rates, accuracy and recall rates marginally exceed 60\%, while precision and F1 scores remain around 20\%. This suggests the presence of numerous random exit events unrelated to QoS metrics in the dataset, preventing the model from learning appropriate predictors. When considering only datasets associated with relevant events (i.e., bitrate switches and stalls), accuracy improves to approximately 90\%, but precision remains around 40\%, indicating a high false-positive rate where many watch continuations are misclassified as exits. However, when utilizing only the dataset containing stall events, all metrics exceed 95\%, demonstrating that using stalls as the dataset significantly reduces interference from QoS-unrelated factors, thereby generating predictors suitable for online system deployment.

\textbf{Sampling Method}
Figure~\ref{img/eva/comparison2_with_f1.pdf} demonstrates the comparison between balanced and unbalanced sampling approaches on the stall dataset. Although model accuracy and precision remain relatively similar, recall decreases by 2\% without balanced sampling, leading to a reduction in the F1 score. This occurs due to class imbalance, where the dominant class (continued watch) dominates the gradient descent process, resulting in some exit instances being misclassified as continued watch, thus increasing the false-negative rate. For a large-scale online system with 400 million daily active users, even a 2\% difference in false-negative rate can potentially cause significant negative system impacts.

\subsection{Pre-deployment Simulation Evaluation}~\label{sec:Simulation Evaluation}
Prior to online deployment, we conduct comprehensive offline evaluations in a simulated environment. The simulation framework incorporates online bandwidth traces and actual video data to ensure environmental fidelity. We select baseline algorithms that optimize the $QoE_{lin}$ objective function (RobustMPC~\cite{yin2015control} and Pensieve~\cite{mao2017neural}), and utilize the video completion rate as the final performance metric. User engagement patterns are modeled through two approaches: deterministic rule-based modeling and data-driven modeling. For RobustMPC, we implement parameter tuning within its $QoE_{lin}$-based decision framework. The Pensieve implementation is augmented to incorporate stall and switching parameters as state variables in its neural architecture, with the reward function dynamically adjusted according to $QoE_{lin}$ parameters during the training phase. This architectural modification enables real-time parameter tuning during inference to accommodate diverse optimization objectives. We plot simulation experiment results as Figure~\ref{The Simulation Experiment of LingXi}, where error bars represent standard errors, and the shaded regions indicate the range of all results with fixed parameters.
\begin{figure*}
    \centering
      \subfigure[Overall Watch time]{
        \includegraphics[width=0.31\linewidth]{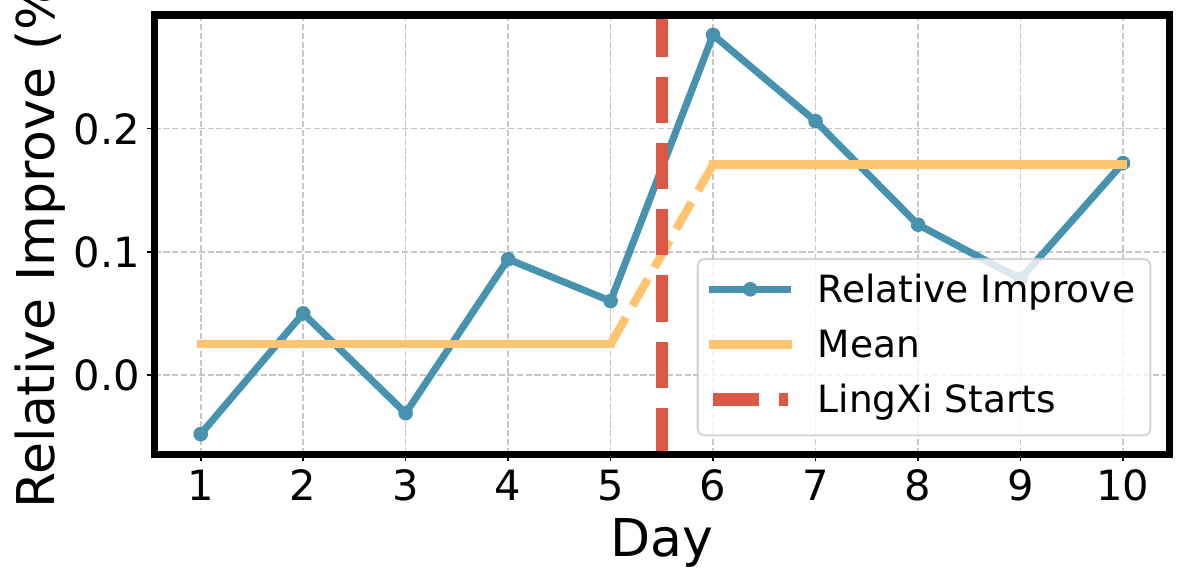}
        \label{Overall Watch time]}
        }
        \subfigure[Bitrate]{
        \includegraphics[width=0.31\linewidth]{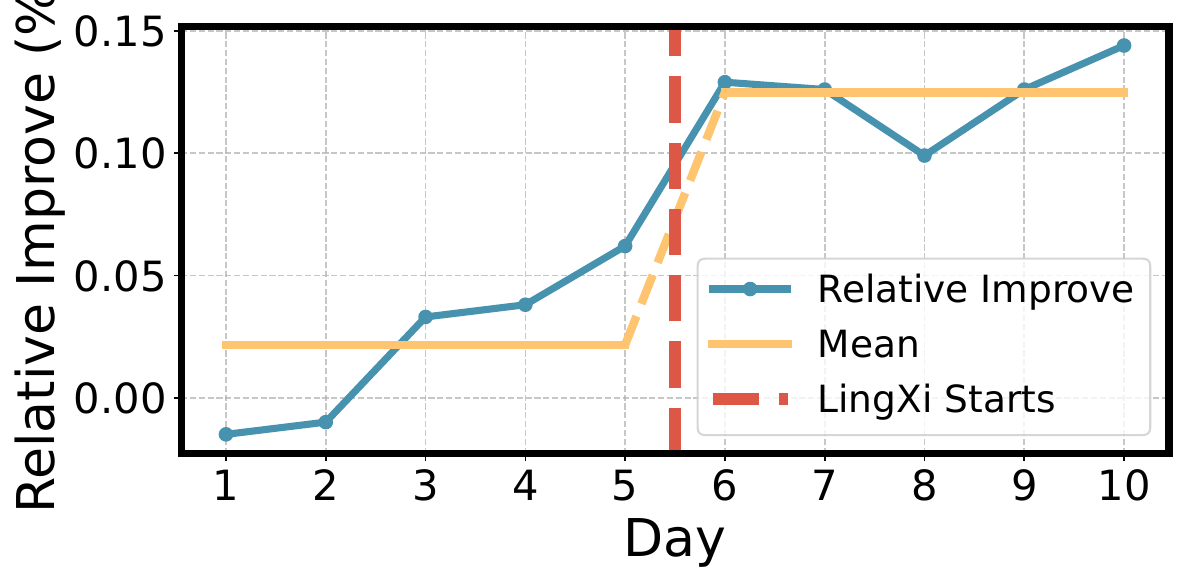}
        \label{Bitrate}
        }
     \subfigure[Stall Time]{
        \includegraphics[width=0.31\linewidth]{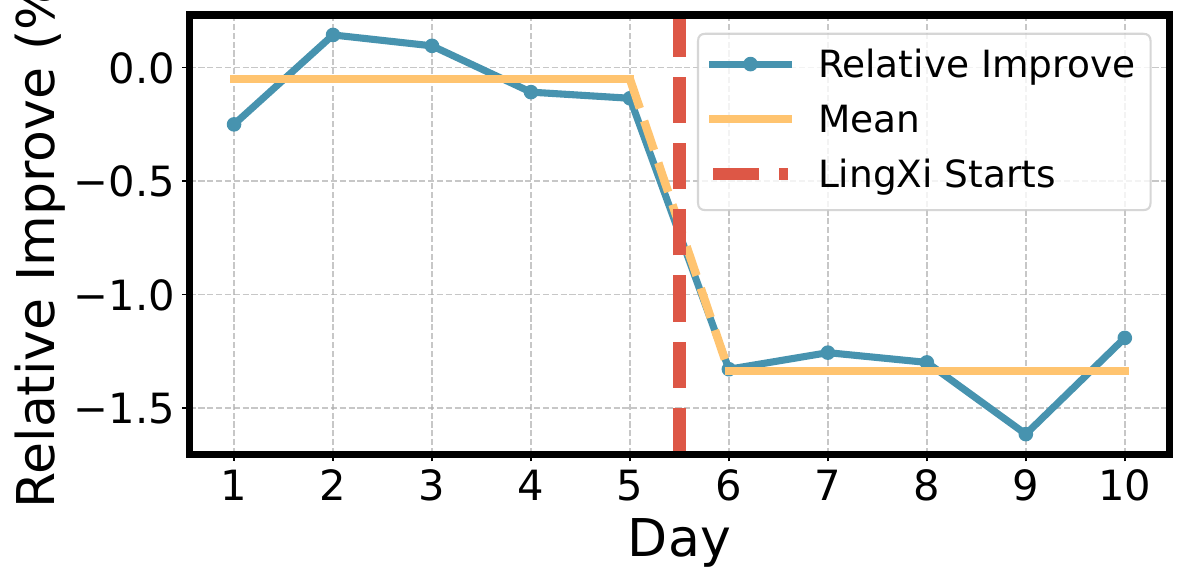}
        \label{Stall Time}
        }
      \vspace{-5pt}
                \caption{The A/B Experiment of \texttt{LingXi}}
        \label{The A/B Experiment of LingXi}
       \vspace{-5pt}
\end{figure*}

\textbf{Rule-Based Modeling}  
The rule-based modeling implements deterministic exit rules based on stall event characteristics. The model incorporates two primary dimensions: cumulative stall time and stall counts. Exit thresholds for both metrics are systematically varied between 2 and 9, generating a comprehensive set of 64 distinct engagement rules. The parametric space for $QoE_{lin}$ encompasses stall parameters ranging from 1 to 20 and switching parameters from 0 to 4.  We compare three methods: fixed parameters, \texttt{LingXi} with fixed candidate parameters~($L(F)$), and \texttt{LingXi} optimized through Bayesian optimization~($L(B)$).

Utilizing the RobustMPC algorithm as the baseline as Figure~\ref{Fixed User, MPC}, the impact of different fixed $QoE_{lin}$ parameters on the completion video rate is minimal. Across various network traces in the dataset, the completion rate ranges from 3\% to 9\%, with the mean under different parameters ranging from 7.3\% to 7.6\%. In contrast, $L(F)$ achieves a mean completion rate of 9\%, a 18\% improvement over the best fixed parameters. $L(B)$ further increases the mean completion rate to 10.9\%, a 43\% improvement over the best fixed parameters. The results for Pensieve are similar to RobustMPC as Figure~\ref{Fixed User, Pensieve}, though the completion rate varies more significantly under different fixed parameters, ranging from 7.2\% to 8\%. However, $L(F)$ achieves a completion rate of 9.6\%, and $L(B)$ reaches 11.3\%, representing a 20\% and 41\% improvement over the best fixed parameters, respectively.

\textbf{Data-Driven Modeling}  
There remain inherent limitations between modeling user engagement with predefined rules and the complexity of real user engagement, making it insufficient as a foundation for online deployment. Therefore, we collect historical engagement data from online users to fit a more accurate user model. We select 100 active users whose daily watch time ranks in the 90th to 95th percentiles and collect their watch  engagement over two weeks. For each user, we fit an individual exit predictor to serve as the user model.

As shown in Figures~\ref{User Model, MPC} and \ref{User Model, Pensieve}, under this modeling method, user performance fluctuates more significantly across different bandwidth traces, but the improvement brought by \texttt{LingXi} becomes more pronounced. Regardless of whether RobustMPC or Pensieve is used as the baseline algorithm, adjusting fixed $QoE_{lin}$ parameters has little effect on the completion rate. In contrast, $L(F)$ with fixed parameters improves the completion rate by 41\% and 34\%, respectively, while $L(F)$ further increases the completion rate by 1.35 times and 54\%, respectively.

\textbf{Detailed Analysis}  
We further analyze the details of algorithm adjustments. In Figure~\ref{Heatmap of Stall Parameters under Different Sensitivities}, we plot a heatmap of the average stall parameters corresponding to different rules under rule-based user engagement modeling. Darker colors represent smaller stall parameters, indicating lower user sensitivity to stall.
It is observed that, regardless of whether RobustMPC or Pensieve is used as the baseline algorithm, the right side of the heatmap is significantly darker than the left. This indicates that as the exit threshold for user stall events increases, \texttt{LingXi}'s perceived user tolerance for stalls correspondingly increases.

\subsection{Large Scale A/B Test}~\label{sec:A/B Evaluation}

\texttt{LingXi} undergoes evaluation through a 10-day difference-in-differences A/B test in the production environment, utilizing 8\% of total traffic. \texttt{LingXi} demonstrates versatility by supporting both explicit QoE-optimizing ABR algorithms and those with implicit objectives. The HYB~\cite{akhtar2018oboe} algorithm exemplifies the latter category, selecting maximum bitrates while maintaining $d_k(Q_k)/C_k < \beta * B$ to prevent stalls. Rather than explicit QoE optimization, HYB employs the $\beta$ parameter to tune algorithmic aggressiveness according to user preferences. We evaluate \texttt{LingXi}'s effectiveness by integrating it with HYB and comparing dynamically optimized $\beta$ values against production-validated static configurations. During the first 5 days, we carried out an AA test to measure the baseline differences between the experimental and control groups without deploying \texttt{LingXi}. On the 6th day, we applied the intervention, enabling \texttt{LingXi} to adjust the $\beta$ parameter, and conducted an AB test comparing \texttt{LingXi} with the control group. The experimental results are shown in Figure~\ref{The A/B Experiment of LingXi}.
\begin{figure}
    \centering
      \subfigure[Online Parameters]{
        \includegraphics[width=0.96\linewidth]{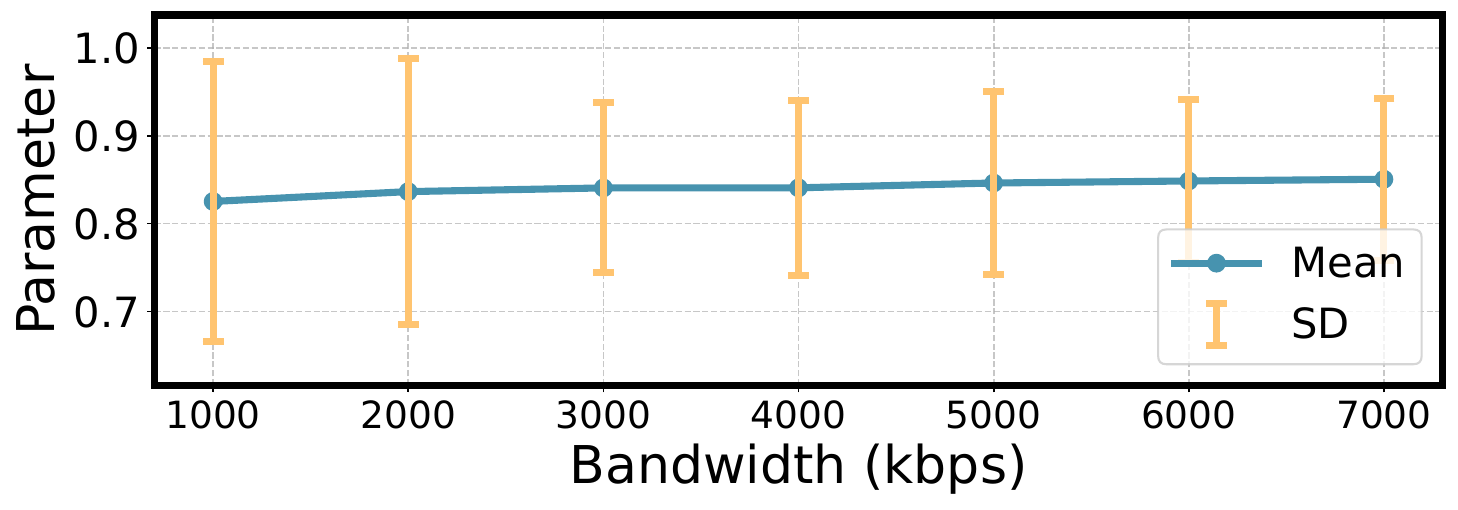}
        \label{Bitrate_difference}
        }
        \subfigure[Relative change in stall time compared to baseline algorithms]{
        \includegraphics[width=0.96\linewidth]{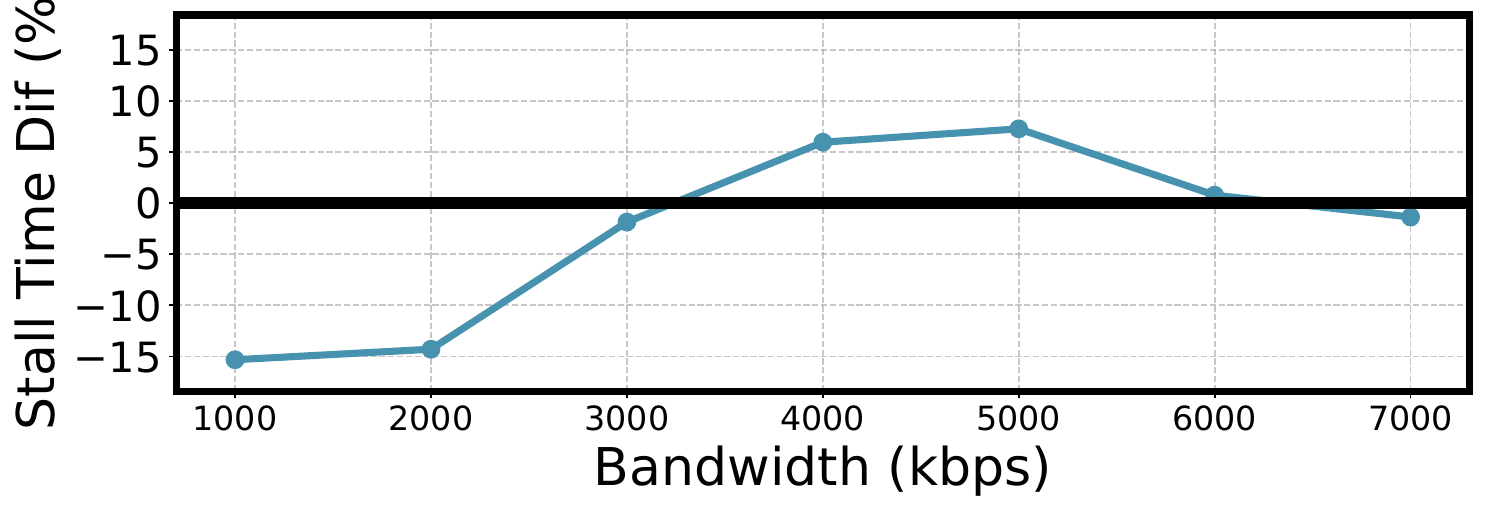}
        \label{rebuf_time_difference}
        }
   \vspace{-5pt}
                \caption{\texttt{LingXi} Performance under Different BW}
        \label{LingXi Performance under Different Bandwidths}
\end{figure}
\begin{figure*}
    \centering
     \subfigure[Day 1]{
        \includegraphics[width=0.32\linewidth]{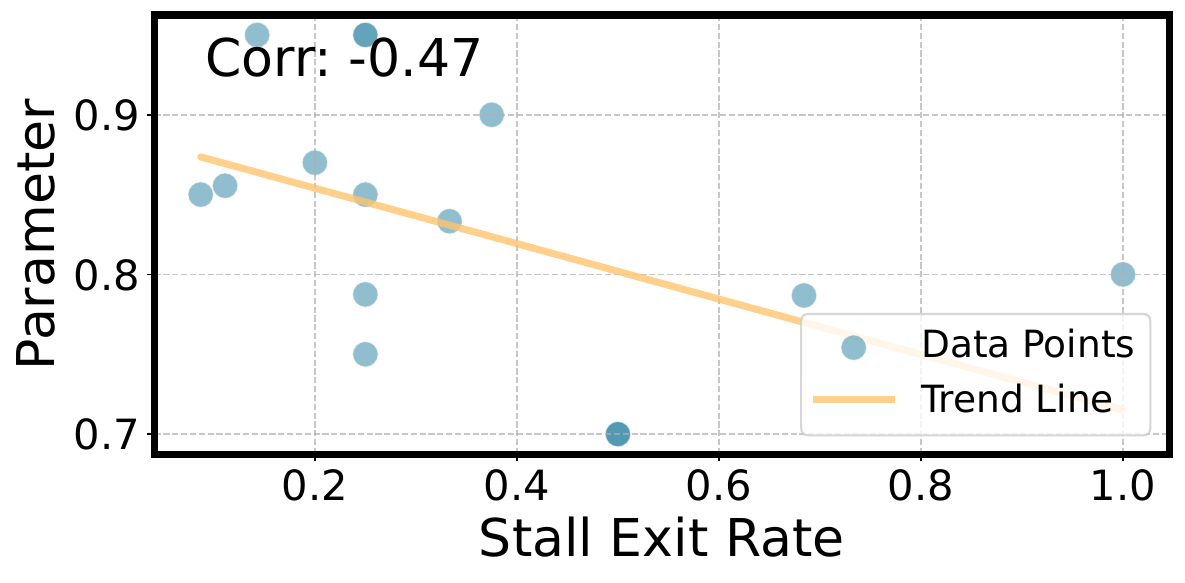}
        \label{Day 1}
        }
        \subfigure[Day 2]{
        \includegraphics[width=0.32\linewidth]{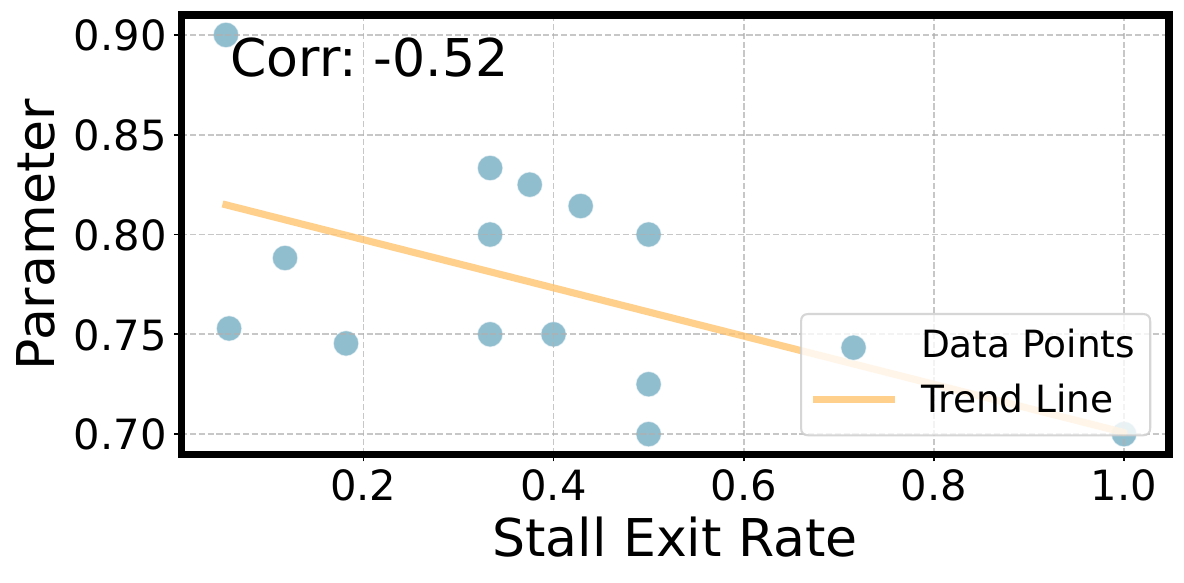}
        \label{Day 2}
        }
         \subfigure[Day 3]{
        \includegraphics[width=0.32\linewidth]{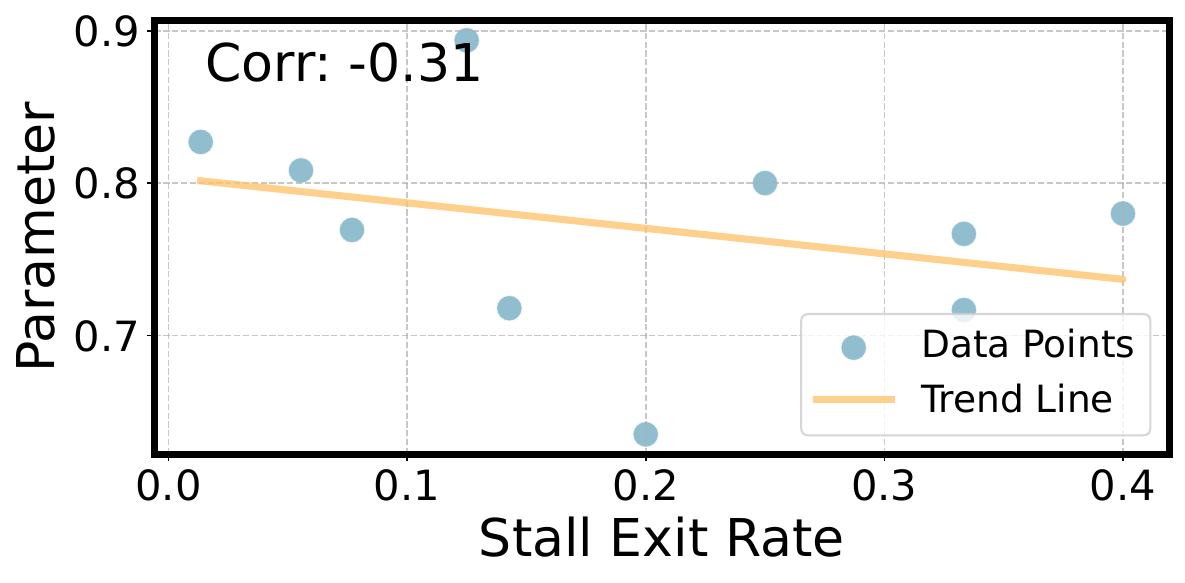}
        \label{Day 3}
        }
         \subfigure[Day 4]{
        \includegraphics[width=0.32\linewidth]{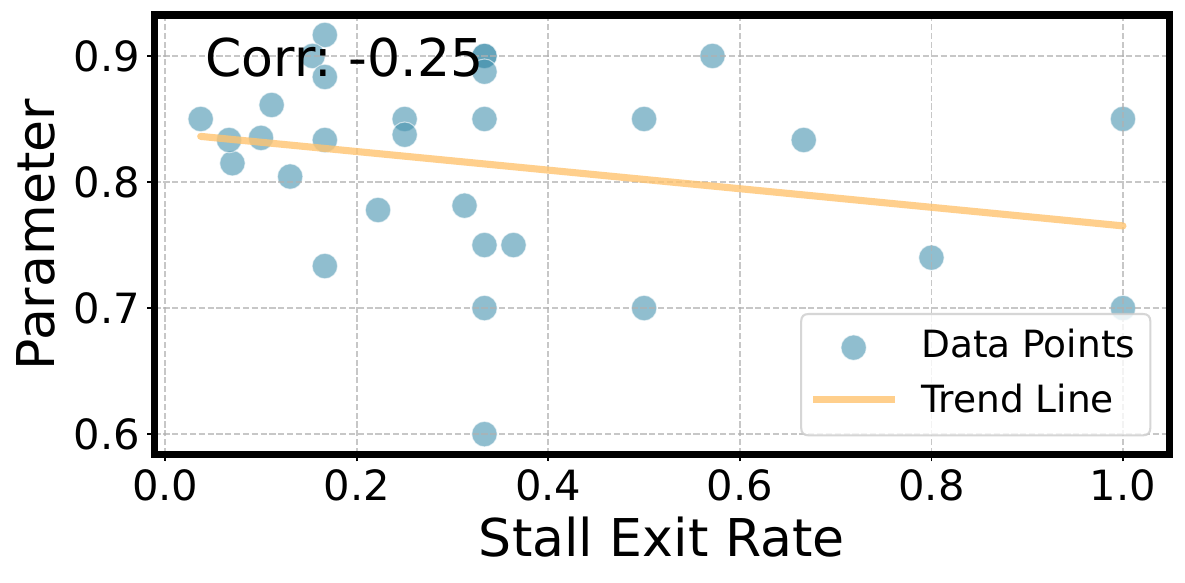}
        \label{Day 7}
        }
      \subfigure[Day 5]{
        \includegraphics[width=0.32\linewidth]{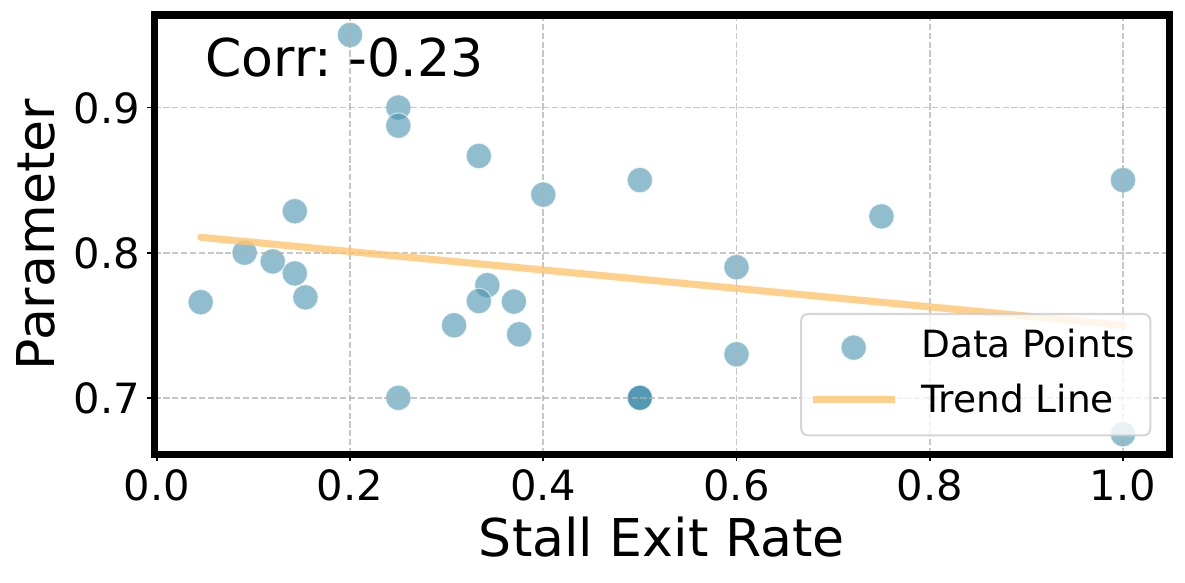}
        \label{Day 8}
        }
          \subfigure[Day 6]{
        \includegraphics[width=0.32\linewidth]{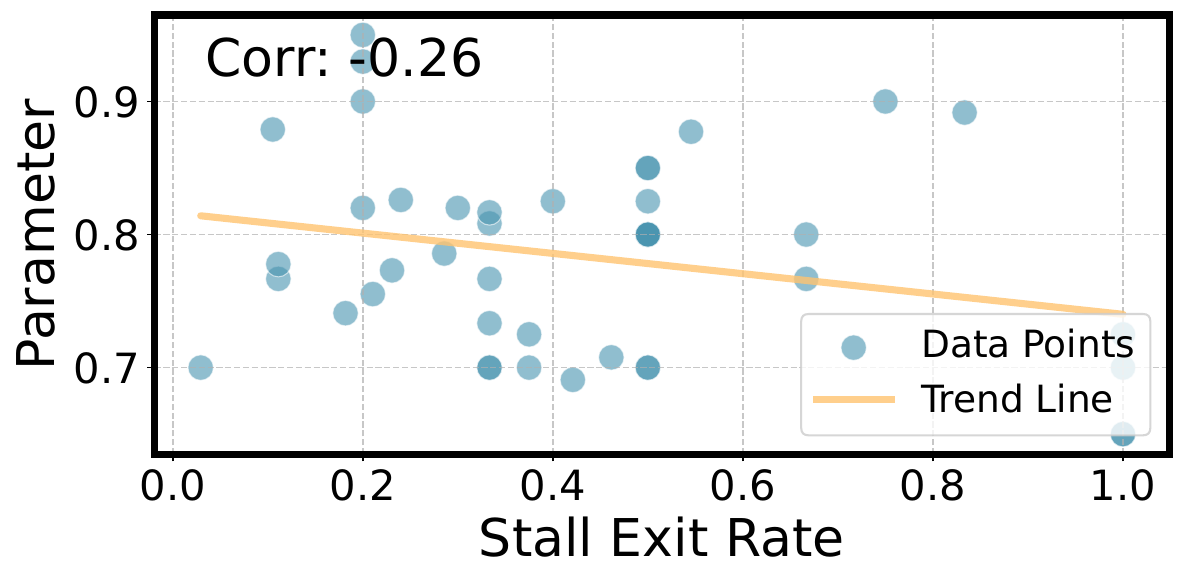}
        \label{Day 8}
        }

         \vspace{-5pt}
        \caption{The Relationship between Stall Exit Rate and ABR Parameter}
        \label{fig: The Relationship between Stall Quit Ratio and ABR Coefficient}
       \vspace{-5pt}
\end{figure*}
\begin{figure*}
    \centering
 
         \subfigure[User 1 (High tolerance)]{
        \includegraphics[width=0.6\linewidth]{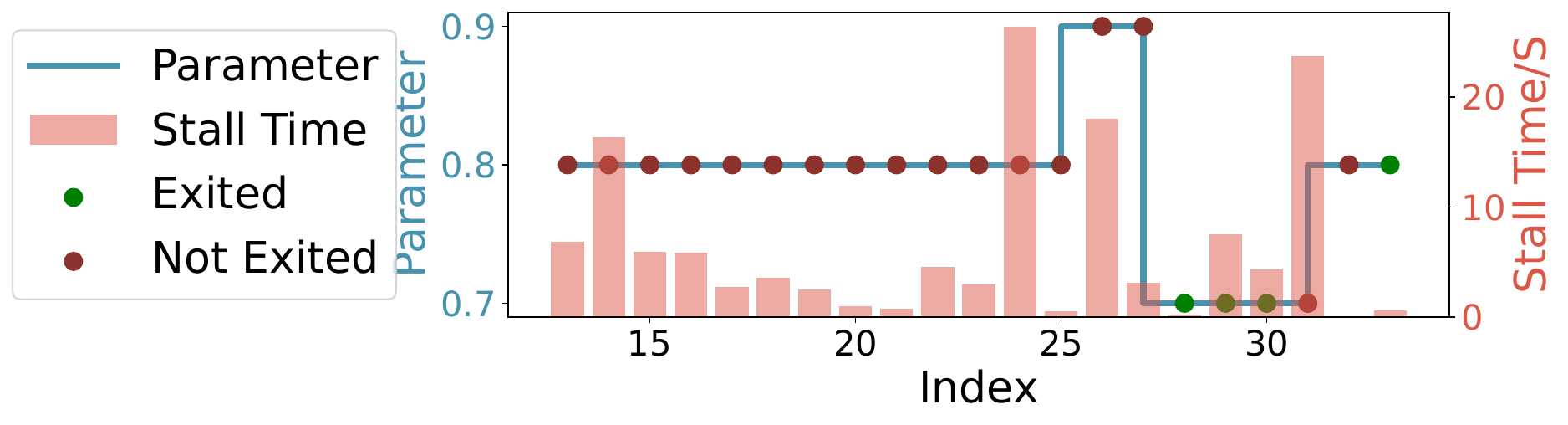}
        \label{User 1}
        }
              \subfigure[User 2 (High tolerance)]{
        \includegraphics[width=0.375\linewidth]{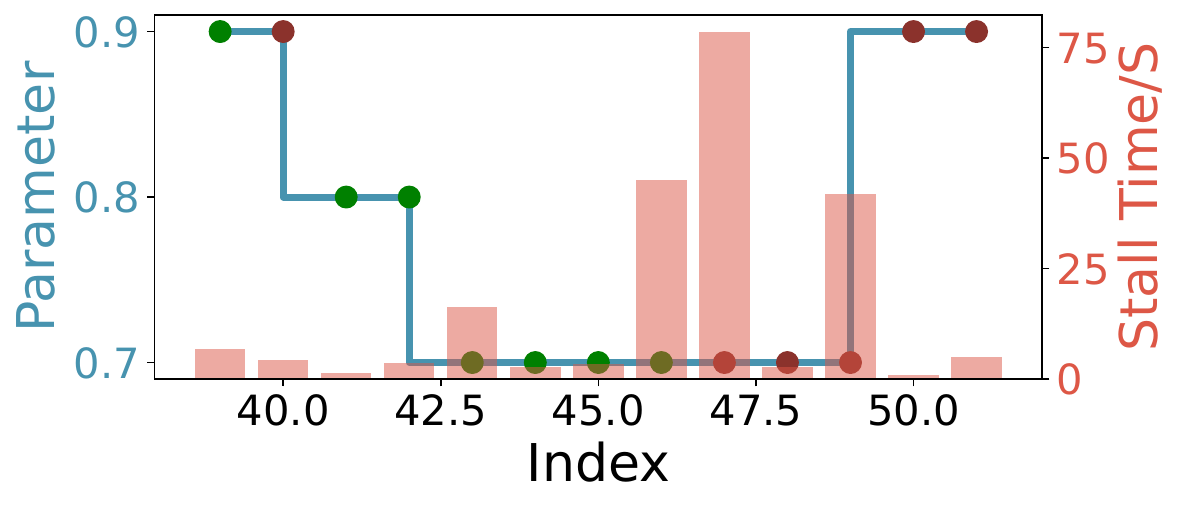}
        \label{User 2}
        }      
         \subfigure[User 3 (Stall-sensitive)]{
        \includegraphics[width=0.6\linewidth]{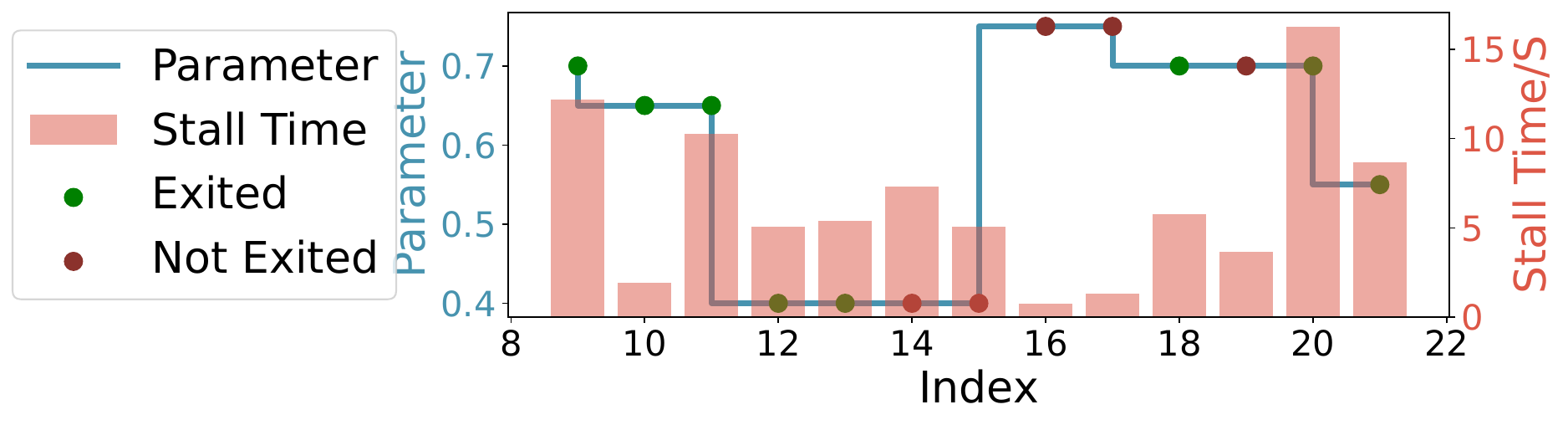}
        \label{User 3}
        }
        \subfigure[User 4 (Stall-sensitive)]{
        \includegraphics[width=0.375\linewidth]{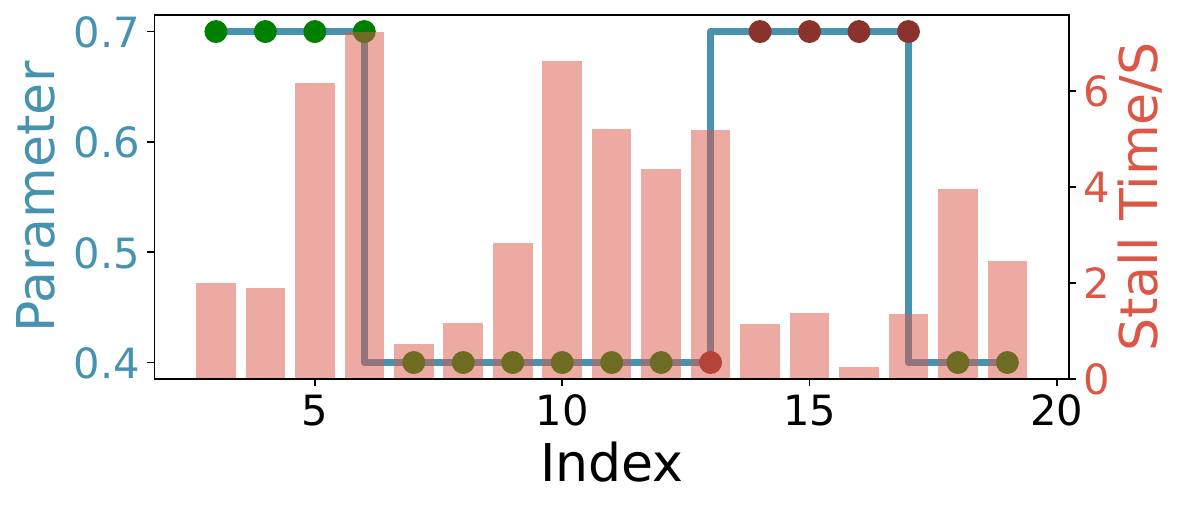}
        \label{User 4}
        }
               \vspace{-5pt}
        \caption{The Details of User Updates to the  ABR Parameter}
        \label{fig: The Details of User Updates to the  ABR Coefficient}
\end{figure*}
\subsubsection{QoE}
QoE serves as \texttt{LingXi}'s primary optimization objective, evaluated through total watch time—the most critical industry indicator. During the pre-intervention period (Days 1-5), the control and experimental groups exhibited comparable baseline performance, with QoE variations contained within ±0.06\%. Following \texttt{LingXi}'s deployment (Days 6-10), the experimental group demonstrated consistent performance gains, with daily improvements ranging from 0.078\% to 0.276\%. The system achieved peak improvement of 0.276\% on deployment day, maintaining robust performance with two subsequent days exceeding 0.2\%. Difference-in-differences analysis confirms that \texttt{LingXi}'s implementation generated a statistically significant increase of 0.146\% ± 0.043\% in total watch time (t = 3.395, p < 0.01). In practical terms, within our experimental cohort of 30 million daily active users, this improvement equates to the watch time contribution of approximately 45,000 additional active users.
\subsubsection{QoS} Although QoS is not the direct optimization objective of \texttt{LingXi}, we observed statistically significant improvements in both video bitrate and stall time. The video bitrate increased by 0.103\% ± 0.015\% (t = 6.867, p < 0.001), while stall time decreased by 1.287\% ± 0.103\% (t = -12.495, p < 0.001). This can be attributed to \texttt{LingXi} dynamically adjusting the optimization objectives to enhance QoE while also achieving traditional QoS optimization goals such as reducing stall and increasing bitrate. Notably, the reduction in stall far exceeds the improvement in bitrate (approximately 12.5 times larger in magnitude), which aligns with our findings in Sec~\ref{sec:Online Log Analysis for QoS}, where stall has a more significant impact on QoE.

\subsection{Focusing on the Long Tail: Performance Enhancement in Low-Bandwidth Scenarios}~\label{sec:long tail}

Large-scale A/B experimental results demonstrate that \texttt{LingXi} achieves statistically significant improvements in QoE and QoS. However, these macroscopic metrics represent the average performance across tens of millions of user experiences, masking potential performance disparities among different user groups. The performance improvements of \texttt{LingXi} primarily benefit bandwidth-constrained long-tail users who are most sensitive to ABR algorithms, while the significant gains for these users are diluted when averaged with data from numerous high-bandwidth users whose optimization potential is inherently limited. Therefore, we conducted an in-depth analysis using detailed playback logs collected from 1/1000 of online users to examine \texttt{LingXi}'s adaptive parameter adjustment process and the stall reduction effects under different bandwidth conditions, aiming to demonstrate that \texttt{LingXi}'s core value lies in providing significant experience improvements for the most critical long-tail users.

\textbf{Parameter Adaptability and Robustness}
In the HYB framework, parameter $\beta$ is crucial for achieving the trade-off between bandwidth estimation confidence and stall risk tolerance. Figure~\ref{Bitrate_difference} shows a positive correlation between $\beta$ values and bandwidth. Specifically, in bandwidth-constrained scenarios where the system faces higher stall risks, \texttt{LingXi} automatically reduces the $\beta$ value, adopting a more conservative strategy while maintaining rapid response to bandwidth fluctuations, resulting in significant $\beta$ variations. Conversely, in high-bandwidth scenarios when stall risk becomes negligible, the system confidently maintains stable and higher $\beta$ values, shifting the optimization objective toward maximizing bitrate to enhance user QoE.

\textbf{Significant Stall Time Reduction}
\texttt{LingXi}'s adaptive strategy ultimately translates into the intuitive performance improvement in stall time reduction. As shown in Figure~\ref{rebuf_time_difference}, optimization effects exhibit significant variations under different bandwidth conditions. In bandwidth-constrained environments below 2000 kbps, \texttt{LingXi}'s advantages are particularly prominent, reducing stall time by up to 15\%, demonstrating its value for long-tail users. As bandwidth conditions improve, stall is no longer the primary concern; therefore, under moderate bandwidth conditions, \texttt{LingXi} strategically shifts optimization focus from stall avoidance to bitrate improvement to enhance video quality. Under these conditions, the similarity in stall rates compared to baseline systems represents an expected strategic trade-off. Finally, in high-bandwidth regions, due to favorable network conditions, stall risks become minimal, optimization space for any algorithm becomes extremely limited, leading to convergent performance across different approaches.

\subsection{Online User Study}~\label{sec:Online User Study}

To further understand how \texttt{LingXi} modifies ABR parameters, we analyzed adjustments from the user's perspective.

\subsubsection{Relationship Between stall Exit Rate and Parameter} In Figure~\ref{fig: The Relationship between Stall Quit Ratio and ABR Coefficient}, we plot the relationship between daily stall exit rates and the $\beta$ parameter. Stall exit rates measure the proportion of stall events that lead to user exit at the current or next video segment. Each data point represents a user's stall-induced exit rate and corresponding parameter value, with trend lines fitted using least squares linear regression based on daily observations. To ensure statistical reliability, the stall-induced exit rates are computed only for users who experience more than 10 stall events per day.

Statistical analysis reveals a robust negative correlation (Pearson correlation coefficient~\cite{spearman1961proof} range: -0.23 to -0.52) between these metrics. Users exhibiting higher stall exit rates are systematically assigned lower $\beta$ values, indicating reduced bandwidth prediction confidence and heightened stall risk aversion. This adaptive mechanism validates \texttt{LingXi}'s capability to quantify individual user sensitivity to stall and dynamically optimize ABR parameters for personalized QoE delivery.
\subsubsection{User-level Analysis}
To examine the dynamic parameter adjustment mechanism, we conducted granular analyses of individual user trajectories. Figure~\ref{fig: The Details of User Updates to the ABR Coefficient} presents temporal adjustment patterns for four representative users, documenting stall events (x-axis), stall time (red bars), parameter evolution (blue line), and user responses (exits in red, continuations in green), respectively.

\noindent\textbf{Stall Sensitivity Detection}
The system exhibits sophisticated sensitivity detection capabilities. User 1's initial high stall tolerance prompted parameter elevation to 0.9, followed by rapid adjustment to 0.7 after consecutive exits during brief stalls (events 28-30). Subsequent parameter restoration to 0.8 following extended stall tolerance (events 31-32) demonstrates an adaptive response to complex sensitivity patterns. Similarly, User 4's parameter trajectory—decreasing from 0.7 to 0.4 during frequent exits, then recovering after sustained viewing through longer stalls—illustrates rapid behavioral adaptation.

\noindent\textbf{User Classification Efficiency}
The system demonstrates robust classification based on stall tolerance patterns. High-tolerance users (Users 1 and 2) maintain parameters within 0.7-0.9, despite temporal variations. User 2's trajectory exemplifies systematic parameter adjustment: initial parameter reduction to 0.7 followed by elevation to 0.9 after consistently demonstrating extended stall tolerance beyond event 47. Conversely, stall-sensitive users (Users 3 and 4) stabilize within 0.4-0.75, with User 3's trajectory showing refined classification through consistent exit patterns and subsequent parameter optimization.

\noindent\textbf{Long-term Stability}
While accommodating short-term behavioral variations, \texttt{LingXi} exhibits convergence toward stable parameter ranges aligned with fundamental user characteristics. This stability manifests in User 1-2's trajectory, maintaining higher parameters despite temporary reductions, and in Users 3-4's consistent lower-range convergence reflecting their inherent stall sensitivity. This equilibrium ensures service stable while preserving adaptation capabilities for necessary adjustments.
\section{Discussion}

\textbf{Extension to Content-based Features}
While \texttt{LingXi} primarily models QoE through application-layer network information, player states, and user engagement history, content-based QoE analysis has emerged as a promising research direction~\cite{zhang2021sensei,dobrian2013understanding}. Network-based and content-based QoE modeling represent two orthogonal approaches, and \texttt{LingXi}'s framework is inherently compatible with content-aware methods. Incorporating content-specific features poses significant deployment challenges. The extraction and processing of content features require complex neural architectures, introducing substantial computational overhead and latency issues in deployment scenarios~\cite{wan2014deep,zolfaghari2018eco}. Our current implementation focuses on network parameters purely from practical deployment considerations. Future work will explore efficient methods to integrate content characteristics and user interaction patterns for enhanced personalized QoE modeling.

\noindent\textbf{Design Choice of \texttt{LingXi}}
\texttt{LingXi} is designed as an optimization objective adjustment module for existing ABR algorithms rather than functioning as a standalone ABR algorithm, based on two key considerations. First, \texttt{LingXi}'s overhead is primarily determined by personalized predictor invocations, which typically consume hundreds of times more computational resources than conventional ABR decisions. Second, directly employing the personalized exit rate predictor 
 as an ABR algorithm would attempt to directly output optimal parameters, which is vulnerable to inherent behavioral uncertainties and complex long-term QoS-QoE interactions~\cite{seufert2014survey}.  Adjusting optimization objectives provides a more reliable and system-safe approach, as it allows the underlying ABR algorithms to adapt while maintaining their built-in stability mechanisms. Therefore, we adopt a Monte Carlo approach in a simulated environment to iteratively adjust ABR optimization objectives, ensuring alignment with user watching patterns. Through carefully designed thresholds, daily active users require at most 1-2 \texttt{LingXi} invocations per day, maintaining system overhead within acceptable limits.
\section{Related Work}

\textbf{QoE in Streaming Systems.}
As an important metric in video streaming systems, QoE has traditionally been assessed subjectively using MOS~\cite{duanmu2018sqoe3}. Modern assessment methods include QoS-based metrics (video quality, stall events) mapped through linear formulas or neural networks~\cite{yin2015control,sqoe4,huang2023optimizing,zhang2023quty}, and engagement-based metrics (watch time, exit rate)\cite{dobrian2011understanding}. While engagement metrics better reflect user experience, their longer statistical intervals make them more popular for CDN-level~\cite{wang2019intelligent,krishnan2012video}.

\noindent\textbf{Personalized Optimization in Streaming Systems.}

\noindent\textit{System-level Personalization} CDN optimization represents a prominent example of system-level personalization~\cite{peng2004cdn}, balancing bandwidth costs and service quality through edge node caching of popular content. It dynamically routes user requests to nodes based on user preferences for latency and video quality~\cite{dobrian2011understanding,krishnan2012video,wang2019intelligent,cheng2023rebuffering,zhang2022aggcast,zhang2020leveraging}.

\noindent\textit{User-level Personalization} User-level personalization primarily remains in small-scale laboratory phases. It mainly models user preferences for different QoS metrics through subjective ratings~\cite{zuo2022adaptive,qiao2020beyond} or playback behavior intervention~\cite{zhang2022enabling}. However, these methods either incur high implementation costs or impact user experience, making deployment in production environments challenging.

\noindent\textbf{Adaptive Bitrate Algorithms.}
ABR algorithms can be categorized into two approaches based on whether they have explicit optimization objectives:

\noindent\textit{Implicit QoE:} Early algorithms primarily focused on single metrics, including throughput-based approaches such as FESTIVE~\cite{jiang2012FESTIVE} and PANDA~\cite{li2014panda}, as well as buffer-based methods such as BBA~\cite{huang2014buffer} and Bola~\cite{spiteri2020bola}. Alternatively, algorithms like HYB~\cite{akhtar2018oboe} select the highest bitrate that avoids stalls based on both throughput and buffer considerations.

\noindent\textit{Explicit QoE:} These directly optimize predefined QoE objectives. Traditional methods include model predictive control MPC~\cite{yin2015control}, online parameter adjustment Oboe~\cite{akhtar2018oboe}, and continuous time modeling SODA~\cite{chen2024soda}. Deep learning approaches include Pensieve~\cite{mao2017neural} based on deep reinforcement learning, Fugu~\cite{yan2020fugu} which uses neural networks to model download duration, Causalsim~\cite{alomar2022causalsim} based on counterfactual reasoning, and Backwave~\cite{jia2024dancing} which employs offline reinforcement learning to address the sim-real gap.

Regardless of whether optimization objectives are explicitly defined, \texttt{LingXi}, functioning as an independent plugin, maintains the capability to adjust parameters to align with user-level QoE.
\section{Conclusion}
We presented the first large-scale deployed user-level personalized adaptive video streaming system, \texttt{LingXi}. The framework continuously monitors user engagement patterns during playback and dynamically optimizes streaming parameters through three key components: a personalized exit rate predictor, Monte Carlo sampling and online Bayesian optimization. Extensive A/B testing in production environments validated \texttt{LingXi}'s effectiveness both in QoE and QoS.

\noindent\textbf{Acknowledgment} We thank the anonymous SIGCOMM reviewers and our shepherd Maria Gorlatova for their constructive feedback. We thank Yangchao Zhao and other colleagues from the Kuaishou Transmission Team for their support in system deployment. This work was supported by the Beijing National Research Center for Information Science and Technology under Grant No. BNR2023TD03005-2, NSERC Discovery grant, Beijing Key Lab of Networked Multimedia, and the Kuaishou-Tsinghua Joint Project.

\noindent\textbf{This work does not raise any ethical issues}

\bibliographystyle{ACM-Reference-Format}
\balance
\bibliography{ref}

\end{document}